\documentclass[letterpaper,preprint,12pt]{aastex}
\pdfoutput=1
\usepackage{amsmath}
\usepackage[width=6.5in,height=9in,top=1in,left=1in]{geometry}

\newcommand\afrho{$A(\theta)f\rho$}
\newcommand\mjysr{MJy~sr$^{-1}$}
\newcommand\invrho{$\rho^{-1}$}
\newcommand\kms{km~s$^{-1}$}
\newcommand\gcm{g~cm$^{-3}$}
\newcommand\inv{$^{-1}$}
\newcommand\sst{\textit{Spitzer Space Telescope}}
\newcommand\spitzer{\textit{Spitzer}}

\newcommand\iras{\textit{IRAS}}
\newcommand\stardust{\textit{Stardust}}
\newcommand\rosetta{\textit{Rosetta}}

\begin{document}

\title{The Dust Trail of Comet 67P/Churyumov-Gerasimenko}
\author{Michael S. Kelley\altaffilmark{a,1,*}, William T. Reach\altaffilmark{b},
  and David J. Lien\altaffilmark{c}}

\affil{\altaffilmark{a}Department of Astronomy, University of
  Minnesota, 116 Church St SE, Minneapolis, MN, 55455}
\affil{\altaffilmark{*} Corresponding author e-mail address:
  msk@physics.ucf.edu}
\affil{\altaffilmark{b}\textit{Spitzer} Science Center, MS 220-6,
  California Institute of Technology, Pasadena, CA 91125}
\affil{\altaffilmark{c}Planetary Science Institute, 1700 E. Ft. Lowell
  Rd. Suite 106, Tucson, AZ 85719}

\altaffiltext{1}{Current address: Department of Physics, University of Central Florida,
  4000 Central Florida Blvd., Orlando, FL, 32816-2385}

\affil{\rm\textbf{Accepted for publication in Icarus}}

\begin{abstract}
We report the detection of comet 67P/Churyumov-Gerasimenko's dust
trail and nucleus in 24~\micron{} \textit{Spitzer Space Telescope}
images taken February 2004.  The dust trail is not found in optical
Palomar images taken June 2003.  Both the optical and infrared images
show a distinct neck-line tail structure, offset from the projected
orbit of the comet.  We compare our observations to simulated images
using a Monte Carlo approach and a dynamical model for comet dust.  We
estimate the trail to be at least one orbit old (6.6 years) and
consist of particles of size $\gtrsim100$~\micron.  The neck-line is
composed of similar sized particles, but younger in age.  Together,
our observations and simulations suggest grains 100~\micron{} and
larger in size dominate the total mass ejected from the comet.  The
radiometric effective radius of the nucleus is $1.87 \pm 0.08$~km,
derived from the \textit{Spitzer} observation.  The \rosetta{}
spacecraft is expected to arrive at and orbit this comet in 2014.
Assuming the trail is comprised solely of 1~mm radius grains, we
compute a low probability ($\sim10^{-3}$) of a trail grain impacting
with \rosetta{} during approach and orbit insertion.
\end{abstract}


\section{Introduction}
Comet dust trails consist of grains that are ejected from comet nuclei
with low velocities and weakly respond to solar radiation pressure.
In contrast, dust \textit{tails} have greater ejection velocities and
are strongly influenced by solar radiation.  These facts have led
investigators to conclude that dust trails are composed of large
grains, in excess of 100~\micron{} \citep[e.g.,][]{sykes92}.  Trails
are so-named because they commonly appear to follow the nucleus along
the comet's projected orbit.  Dust trails are long lived
($\lesssim100$~yr), a result of normal comet activity, and are the
principal mass-loss mechanism of short-period comets, outside of
fragmentation and total disruption \citep{sykes92, reach00, reach07}.
Over time, planetary perturbations displace a comet's trail from the
nucleus and the trail becomes increasingly tenuous.  The trail is then
considered to be a meteoroid stream and produces meteor showers if the
stream and a planet collide (e.g., the meteor showers associated with
comets 2P/Encke and 55P/Tempel-Tuttle).  Space-based mid-infrared
(mid-IR) observations are readily sensitive to thermal emission from
trail particles and the detection rate of trails in Jupiter-family
comets is $>80$\% \citep{reach07}, but optical observations of trails
from the ground are possible and have been presented for comets
2P/Encke, 22P/Kopff, and 81P/Wild \citep{ishiguro02, ishiguro03,
  ishiguro07}.

The total mass of a dust trail can be estimated from images and
dynamic models, assuming the structure and density of the trail dust
grains.  Comet mass-loss estimates that include trail/large particle
production infer dust-to-gas mass ratios that are larger than 1 and
that comet nuclei are appropriately described by \citet{sykes92} as
``icy mud balls.''  For example, \citet{reach00} derived a shallow
grain size distribution for comet Encke from dynamical simulations of
the comet coma and trail at $r_h=1.2$~AU.  Their grain size
distribution ranges from $dn/da \propto a^{-0.7}$ to $a^{-0.4}$,
yielding a dust-to-gas mass ratio of $\approx 10$--30.  The large
derived ratio is mostly due to the dynamics of the coma, best
described by large, fast moving particles.  \citet{sykes92} deduce a
dust-to-gas mass ratio of 3.5 from analysis of Encke's dust trail and
\citet{lisse04} posit a value of 2.3 with \textit{Infrared Space
  Observatory} photometry of the coma.  In contrast, comet Encke's
dust-to-gas mass ratio derived from optical observations suggest the
ratio is an order of magnitude lower \citep[e.g., 0.2;][]{osip92}.
The large dust output from comets imply that comet dust trails are a
significant input to the interplanetary dust complex \citep[about half
  of the dust required to replenish the zodiacal dust complex inside
  of 1~AU;][]{sykes04}.

In the following paper, we present a study of the dust trail of the
ecliptic comet 67P/Churyumov-Gerasimenko (67P).  Comet 67P is the
primary mission target of the European Space Agency's \rosetta{}
spacecraft, designed to characterize the comet nucleus (morphology,
composition) and the comet coma (development of activity, dust-gas
interaction, interaction with the solar wind) by following, orbiting,
and landing a probe on the nucleus \citep{glassmeier07}.
Characterization of the comet's gas and dust environment is important
to mission planning (see \citealt{colangeli04} and references therein;
\citealt{agarwal07b}).  \citet{fulle04} identified this comet to have
a neck-line tail structure in optical observations of the comet and
concluded that the comet is significantly active dust at 3.6~AU
pre-perihelion; \rosetta{} will enter orbit around the comet at
4--5~AU pre-perihelion.  \citet{sykes92} have shown 67P to have dust
trail in observations of the comet by the \textit{Infrared
  Astronomical Satellite} (\iras) and estimated this comet's total
dust-to-gas mass ratio to be 4.6.  When the trail was observed after
the 1982 perihelion, the trail spanned from 0.1\degr{} in mean anomaly
ahead to 1.1\degr{} behind the comet nucleus.  \citet{agarwal07a} have
detected the dust trail at optical wavelengths with the MPG/ESO 2.2~m
telescope (4.7~AU post-perihelion).  In this paper, we report the
detection of 67P's dust trail and nucleus in \sst{} images taken at
4.5~AU, post-perihelion.  We discuss the comet nucleus and dust
observed in the mid-IR images and compare our observation to Palomar
Observatory ground-based optical images in \S\ref{sec:results}.  We
model the dust environment of 67P with dynamic models
(\S\ref{sec:dynamics}-\ref{mcmodel}) to estimate the trail grain
velocities, sizes, and ages
(\S\ref{sec:modelvej}-\ref{sec:dustnature}).  Finally, we estimate the
impact hazard the trail presents to the \rosetta{} spacecraft
(\S\ref{sec:impact}).  We summarize the paper in
\S\ref{sec:conclusions}.

\section{Observations and Reduction}
\sst{} \citep{gehrz07, werner04} images of 67P were taken on
2004~Feb~23 03:44~UT with the Multi-band Imaging Photometer for
\spitzer{} \citep[MIPS;][]{rieke04} 24~\micron{} camera
($\lambda=23.7$~\micron, $\Delta\lambda = 5.3$~\micron).  The
observation \dataset[ADS/Sa.Spitzer#006612736]{(\spitzer{}
  astronomical observation request key 0006612736)} consists of a
$3\times1$ mosaic of 10~s exposures using the MIPS
photometry/super-resolution observation template for a total
integration time of 140~s per pixel (one-sigma sensitivity of
0.055--0.077~\mjysr{} per pixel).  One targeted pointing of the MIPS
photometry/super-resolution mode consists of fourteen dithered images
(2.5~arcsec~pixel\inv) providing an $8\arcmin\times5$\arcmin{}
field-of-view, centered on the target.  In support of the \spitzer{}
observation, we also imaged comet 67P with the Mount Palomar 5~m Hale
telescope Large Format Camera (LFC, a prime-focus camera and mosaic of
six detectors spanning a 24\arcmin{} diameter field-of-view) on 25--27
June 2003.  Our viewing geometries are summarized in
Table~\ref{table:obsgeom} and a diagram of the Earth, \spitzer{}, and
67P is presented in Fig.~\ref{fig:orbit}.  The \iras{} observation
geometry is included for comparison.  Comet 67P is a Jupiter-family
comet with a 6.6 year period (eccentricity = 0.63, semi-major axis =
3.5~AU, inclination = 7.1\degr) and we observed the comet 312~days
($\lambda=0.6$~\micron{}) and 554~days ($\lambda=24$~\micron{}) after
the August 2002 perihelion.

MIPS images were initially processed with the \spitzer{} Science
Center's pipeline version S11.4.0.  Starting with individual basic
calibrated data (BCD) frames, we applied a few corrections, as
prescribed by the MIPS Data Handbook \citep{mdh}, before creating the
final mosaic.  Our individual frames suffered from what is known as
the ``jail bar effect,'' that is, the readout of one saturated pixel
causes a constant offset in every forth column.  We subtracted a
constant from each affected column such that the median value matches
the median value of the nearby unaffected columns.  We added a
constant value to each frame to correct for slight background offsets
between dither positions.  Next, we median combined all data frames to
create a clean ``delta-flat field.''  We divided all frames by the
delta-flat field to correct for data variations between the data and
the pipeline flat field.  Finally, distortions in the MIPS
24~\micron{} focal plane were removed by mapping the frames onto a
plane tangent to the celestial sphere using the distortion solution
provided by the \spitzer{} Science Center.  The resulting images were
mosaicked, accounting for the motion of the comet, to create the final
image.  The delta-flat field, distortion corrections, and mosaicking
were performed with the \spitzer{} Science Center's MOPEX software
\citep{makovoz05}.  The final mosaicked image, rotated to the
celestial coordinate system, is presented in Fig.~\ref{fig:mips}.

Optical images were obtained on three nights, 25--27 June 2003 UT,
using the Gunn $r^\prime$ filter \citep{fukugita96}.  The total
integration time on the comet was 55~min, observed at an average
airmass of 1.80, and an average seeing of 1.9\arcsec.  The LFC was
placed in $2\times2$ binning mode, for an observed platescale of
0.36~arcsec~pixel\inv.  The images were reduced using standard
techniques and the NOAO IRAF \citep{tody93} Mosaic Reduction Package.
Color differences between twilight/dome light and the dark sky
affected the mosaic flat fielding.  Each night's deep exposure frames
were object masked and median combined to create super-flats that
mitigate the effect of the color difference.  Unaccounted features in
the flat-field and slight offsets in the chip-to-chip background
matching produced the large scale artifacts apparent in the co-added
image.  The structure varies on the order of 0.5\% of the background
and were most apparent in images from the last night of observations
for which a complete flat-field could not be derived.

The Palomar images were photometrically calibrated with the
\citet{smith02} Sloan $r^\prime$ standard stars SA 107-351, and SA
110-232.  A canonical airmass correction of 0.115 magnitudes per
airmass was derived from the sky extinction at Palomar and the
$r^\prime$ bandpass.  The error in standard star photometry was 1.7\%.
Altogether, our signal-to-noise ratio limit was $\approx50$, that is,
the flat-field artifacts and standard star photometry limited our
photometric error to no less than 2\%.  Our three nights of 67P images
were combined in the comet's rest frame to remove background objects,
and the result is presented in Fig.~\ref{fig:lfcimage}.

\section{Results}\label{sec:results}
\subsection{Comet Morphology}
The \spitzer{} image clearly shows emission in both the forward and
backward directions.  Throughout the paper, forward refers to the
direction of the comet's projected velocity vector, and backward
refers to the anti-velocity direction (see Figs.~\ref{fig:mips}
and~\ref{fig:lfcimage}).  In Fig.~\ref{fig:mips}, it is important
to note the asymmetry in the backward direction.  The bottom panel of
Fig.~\ref{fig:mips} plots contours to enhance the comet morphology.
The bulk of the backward emission is found to the south of the
projected orbit, meanwhile the forward emission is centered on the
orbit.  The ephemeris of comet 67P (computed 2004 Aug 23, solution JPL
K023/22) has an observation baseline of 15.96 years and the
root-mean-square of the fit residuals is 0.7\arcsec, or 0.3~pixels.
The astrometry of the field was verified with the 2MASS point source
catalog.  The central point source of 67P is found within 0.4~pixels
of the expected position.  The alignment of the forward emission with
the orbit and the asymmetry of the backward emission with respect to
the orbit appear to be true, given the accuracies of the image and
orbit solution.

The Palomar image is presented in Fig.~\ref{fig:lfcimage} and compared
to the projected orbit at that epoch.  Here the backward emission lies
to the north of the projected orbit and is visible to 10\arcmin{}
behind the nucleus.  With close inspection of the data, dust emission
is evident out to 14\arcmin.  The forward emission observed in the
24~\micron{} image does not appear to be present in the 0.6~\micron{}
image, at least at this epoch and sensitivity limit.  We identify the
natures of the forward and backward emissions through comparisons to
models of dust comae in \S\ref{sec:dustnature}.

\subsection{Comet Nucleus and Dust Production}\label{sec:nuke}

The central point source in the 24~\micron{} image has a full-width at
half maximum (FWHM) of 6.5\arcsec{} and a surrounding diffraction
ring, similar to point sources in the surrounding field.  At the
observed distance of $\Delta=4.1$~AU, the point source width is
19,000~km; such a large area may still contain flux from an unresolved
coma.  This dilemma is typically encountered in observations of comets
at large heliocentric distances.  At best we can consider the point
source to be solely due to emission from the nucleus, and at worst the
point source is dominated by an unresolved coma.  We assume that the
point source flux and the nucleus flux are equivalent but at this
point we note that without more information this estimate is truly an
upper-limit.

The emission from the nucleus was derived by fitting a point-spread
function (PSF) to the comet's central point source with IRAF's DAOPHOT
package.  Upon initial PSF subtraction, it became apparent that the
inner 10\arcsec{} of coma (superposed with the nucleus) could either
be as faint as the forward emission, or as bright as the backward
emission.  The flux of the nucleus could be modified to reproduce
either case, with both cases being equally viable given the
signal-to-noise ratio of our image.  We considered the two cases to be
upper- and lower-limits to the point source flux.  A color-corrected
flux density of $3.1\pm0.2$~mJy ($1\sigma$ formal error) was obtained
by matching the inner coma to the emission ahead of the nucleus, and a
flux density of $2.8\pm0.2$~mJy when the inner coma is matched to the
backward emission.  Together, these values suggest a flux density of
$2.95\pm0.25$~mJy.  A cut across the point source before and after
point source subtraction (at 2.95~mJy) is presented in
Fig.~\ref{fig:pointsource}.  The total flux density inside a 10~pixel
radius aperture before point source subtraction was $17.53 \pm
0.37$~mJy.

To estimate the effective radius of the comet nucleus we use the
near-Earth asteroid thermal model (NEATM) of \citet{harris98} with a
geometric albedo, $p_v$, of 0.04 and an IR emissivity, $\epsilon$, of
0.9.  The NEATM formalism requires the IR-beaming parameter to be fit
to the observed color-temperature.  In the absence of any color
information on the nucleus, we chose the IR-beaming parameter to be
0.756, the same value used for main-belt asteroids \citep{lebofsky86}
observed at phase angles less than 30\degr{} (our observation is at a
phase angle of 12\degr).

The NEATM only applies to slowly rotating objects.  Slow rotator
models assume each point on the surface is in instantaneous
equilibrium with insolation, i.e., the sub-solar point is the hottest
point and the night side is the coldest.  In contrast, a fast rotator
model applies to surfaces that have no time to radiatively cool
through the night side, and, therefore, is isothermal with respect to
solar latitude.  Following \citet{spencer89}, we can test if the 67P
nucleus may be considered a fast or slow rotator at a given
heliocentric distance.  The unit-less parameter $\Theta$ determines
the applicability of the two models,
\begin{equation}
\Theta = \frac{\Gamma \sqrt{\omega}}{\epsilon \sigma T^3_{ss}},
\label{eqn:theta}
\end{equation}
where $\Gamma$ is the thermal inertia in MKS units
(J~K$^{-1}$~m$^{-2}$~s$^{-1/2}$, commonly abbreviated as MKS),
$\omega$ is the angular rotation rate of the object, and $\sigma$ is
the Stefan-Boltzmann constant.  Slow rotators have $\Theta \ll 1$ and
fast rotators have $\Theta \gg 1$.  For the thermal inertia of 67P's
nucleus, we choose an upper-limit of 100~MKS, constrained by the
$\Gamma$ upper-limit of the comet 9P/Tempel nucleus from \textit{Deep
  Impact} fly-by and \sst{} observations \citep{ahearn05, lisse05}.
Lower values are more probable \citep{groussin06}, and $\Gamma
\lesssim 20$~MKS has been measured for Centaurs 95P/Chiron,
(8405)~Asbolus, and (10199)~Chariklo \citep{fernandez02, groussin04}.
The rotation period of the 67P nucleus is $12.3 \pm 0.27$~hr
\citep{lamy04} and the sub-solar temperature $T_{ss} = 204$~K, derived
from our NEATM model.  If the 100~MKS upper-limit to the thermal
inertia of the 67P nucleus is appropriate, then we calculate a
$\Theta$-value of $\lesssim 1.2$, suggesting a slow rotating nucleus,
or, less probably, an intermediate case between a fast and slow
rotating nucleus.  The NEATM is a valid model for the 67P nucleus at
4.5~AU.

We derive an effective radius of $1.87 \pm 0.08$~km for the nucleus.
The error at the $3\sigma$ level ($\pm 0.24$~km) encompasses the full
$3\sigma$ range of point source fluxes, 2.2--3.7~mJy, estimated from
the two PSF fits above.  Our derived effective radius is in agreement
with the radius $1.98 \pm 0.02$~km calculated from \textit{Hubble
  Space Telescope} observations (\citealt{lamy06}; see also
\citealt{lamy07}), also using a geometric albedo of 0.04, although the
infrared derived radius is not as sensitive to the chosen albedo as is
the optically derived radius.  Considering the agreement between
effective radii, there likely is little or no unresolved coma in our
flux estimate for the nucleus.  We can combine the optical and
infrared results to derive the comet's true geometric albedo and find
a value of $0.035\pm0.005$.  The calculation of the albedo assumes the
cross-section of the nucleus at the epoch of the \textit{Hubble} and
\spitzer{} observations are equivalent.  This is not necessarily the
case.  Indeed, the axial ratio of the nucleus derived from the
\textit{Hubble} light curve is $\geq1.55$, i.e., the effective
cross-sectional area varies by at least a factor of 1.55 over one
12~hr period, depending on the orientation of the spheroid.  Since the
relative phase of the \spitzer{} observation is unknown (1.87~km could
be a light-curve maximum, minimum, or anything in-between), the
solutions to the geometric albedo range from 0.015 to 0.078, wholly
encompassing the known comet nucleus albedo range of 0.02--0.06
\citep{lamy04}.  With some confidence we can state that the albedo of
67P is not radically different from typical comet surfaces.

In the Palomar image, the coma has a radially averaged FWHM of
11\arcsec{} (28,000~km) and is asymmetric with respect to the orbit,
with more emission found to the southwest.  If the coma is in a
``steady state'' then we can estimate the product \afrho{}.  The
product \afrho{} is commonly used as a proxy for dust production
\citep{ahearn95}, where $A(\theta)$ is the dust albedo as a function
of $\theta$, the scattering angle, $f$ is the filling factor of the
dust grains in the aperture, and $\rho$ is the radius of the
field-of-view \citep{ahearn84}.  A steady-state coma arises from a
spherically symmetric, homogeneous outflow of dust, resulting in an
observed coma flux that varies linearly with aperture radius.  As a
consequence, the steady-state coma's \afrho{} is independent of radius
\citep{ahearn84}.  The asymmetry in the coma morphology of 67P
prevents us from assuming a steady state coma (here, the integrated
aperture profile varies as $\rho^{1.4}$ instead of $\rho^{1.0}$), but
in order to compare our data to other investigations we continue with
an estimate of \afrho{}.  We derive values of \afrho{} that range from
$10.4\pm0.2$ to $25.8\pm0.4$~cm for apertures of radius
3400--68000~km.

Comet 67P's \afrho{} values at 3.23~AU are approximately a factor of
3--5 lower than extrapolation from the observations of
\citet{ahearn95} using the averaged power-law, \afrho{}$=323
r_h^{-1.34}$~(cm), but the values agree given the error in the slope
($\pm0.81$).  \citet{kidger03} derives \afrho{}$=1530
r_h^{-5.8}$~(cm), which extrapolates to 2--3~cm at $r_h=3.2$~AU.  Our
\afrho{} value is approximately 7 times larger.  The \citet{kidger03}
investigation is a collection of 625 standardized amateur astronomer
CCD observations from perihelion to 2.8~AU post-perihelion (whereas
\citeauthor{ahearn95} measured 13 observations from perihelion to
1.9~AU post-perihelion).  The visual light curve\footnote{Available at
  S.\ Yoshida's Comet Catalog:
  http://www.aerith.net/comet/catalog/0067P/2002.html} varies as
$r_h^{-4}$, therefore, the steeper slope (-5.8) likely indicates the
overall dust production trend.  \citet{schleicher06} measured \afrho{}
and the gas production of comet 67P and found the gaseous species
(such as OH, a proxy for water production) follow steep profiles
($r_h^{-4}$ to $r_h^{-6}$) and \afrho{} to vary as $\sim r_h^{-1}$.
If \afrho{} truly is the dust production, then the shallow profile
implies that as the comet recedes from the Sun, the volatile gases
become increasingly efficient at ejecting dust.  A more probable
scenario, as \citet{schleicher06} points out, is that slow moving,
medium-sized (10--100~\micron) particles are lingering near the
nucleus, causing the shallow slope in the post-perihelion \afrho{}
dependence on $r_h$ (the comet is not as well studied at
pre-perihelion epochs).  The factor of 7 discrepancy between the LFC
derived \afrho{} value at 3.23~AU and the \citet{kidger03}
extrapolated value could be consistent with a lingering population of
medium-sized dust grains.  Alternatively, a power-law description of
the dust production is likely a simplification.  At 3.23~AU
post-perihelion, water sublimation continues to decrease, and other
ices, such as CO and CO$_2$ become increasingly important as drivers
of dust production \citep{meech04} and a power-law does not account
for discrete, yet prolific, dust production by jet activity related to
seasonal temperature changes on the surface and sub-surface.  We adopt
$r_h^{-5.8}$ as an approximation to the dust production trend for the
purposes of dynamical simulations of the coma and trail
(\S\ref{sec:dynamics}).  The dynamical simulations automatically take
into account large, slow moving grains by treating the dynamics of
dust as a function of grain size.

\subsection{Surface Brightness Profiles and Albedo}\label{sec:profiles}
To compare the Palomar and \spitzer{} data, we binned each image with
rectangular apertures (0.6~\micron{}: $22.5\arcsec\times7.5\arcsec$,
24~\micron{}: $21\arcsec\times7.4\arcsec$), where the long dimension
is placed parallel to the orbit.  Figure~\ref{fig:profile} presents
the profiles, de-projected according to the angle listed in
Table~\ref{table:obsgeom}.  The profiles can be approximated by
power-laws and the best fits to the profiles, in terms of surface
brightness ($S_\nu$) and optical depth ($\tau$), are presented in
Table~\ref{table:profilefit}.  The 0.6~\micron{} $\tau$ was calculated
using the solar spectral energy distribution from \citet{neckel84} and
the 24~\micron{} $\tau$ was calculated using the blackbody temperature
at a heliocentric distance of 4.47~AU.  The ratio $\tau_{0.6}$ to
$\tau_{24}$, the albedo of the grains, ranges from 0.05 to 0.10 at
$2\times10^6$ to $0.5\times10^6$~km from the nucleus.  The
24~\micron{} profile is shallower than the 0.6~\micron{} profile by a
factor of $d^{0.5}$, where $d$ is the distance from the nucleus.  The
difference suggests that the images are sampling two disparate grain
populations (either physically or dynamically different) and that a
direct photometric comparison (e.g., the albedo calculation) is not
appropriate.

Figure \ref{fig:perp-profile} is a plot of cuts from the point
source subtracted 24~\micron{} image.  Each cut is perpendicular to
67P's projected orbit.  The emission has an approximate width of
60,000~km ahead of the nucleus and 60,000--120,000~km behind the
nucleus.  As discussed in \S\ref{sec:dustnature}, the peak of the
backward emission is not aligned with the orbit.

\section{Discussion}
\subsection{Dynamical Model}\label{sec:dynamics}
Comet dust is ejected from the nucleus by surface and sub-surface
volatile sublimation.  Once a dust grain decouples from the outward
gas flow in the near-nucleus environment, the principal forces
remaining are solar gravitational and solar radiation forces.  Since
both forces are proportional to $r_h^{-2}$ and the force of solar
radiation ($F_{rad}$) opposes the force of gravity ($F_g$), the net
effect may be considered a reduced solar gravitational force.  This
effect is commonly parameterized by $\beta$, the ratio of the
radiation force ($F_{rad}$) to the gravitational force ($F_g$),
\begin{equation}
F_{net} = F_{rad} + F_g = (1 - \beta) F_g,
\label{eqn:fnet}
\end{equation}
where $\beta$ reduces to
\begin{equation}
\beta = \frac{0.57 Q_{pr}}{a\rho}, \label{eqn:beta}
\end{equation}
and $Q_{pr}$ is the efficiency of radiation pressure on the grain, $a$
is the radius of the grain in units of \micron, and $\rho$ is the
grain density in g~cm$^{-3}$ \citep{burns79}.  The $\beta$
parameterization is used throughout this paper.  To approximate the
more rigorous treatment to grain structure in thermal emission and
light scattering models of comet dust, the dynamic model uses a low
dust grain density equal to 1~g~cm$^{-3}$ for materials with bulk
densities ranging 2--3~g~cm$^{-3}$ \citep[cf.,][]{lisse98, harker02,
  kimura06}.  The model also assumes $Q_{pr}=1$.  This value is
appropriate for large, isotropic scatterers with $Q_{abs} \approx 1$
\citep{burns79}, i.e., the modeled grains are large with respect to
the absorbed light (here, the solar spectrum).  To remain within the
large particle limit, our simulations only choose particles with
$a\gtrsim0.5$~\micron{}.

Comet grain densities lower than our chosen value of 1~g~cm$^{-3}$
have been deduced by other investigations.  For example:
\begin{enumerate}
\item \citet{fulle00} use \textit{in situ} observations and dynamic
  models to constrain comet 1P/Halley's dust grains to densities
  ranging from 0.05--0.5~g~cm$^{-3}$, with 0.1~g~cm$^{-3}$ being the
  favored solution.

\item \citet{lasue06} model the dust of comet C/1995~O1 (Hale-Bopp)
  using aggregates and compact spheroidal grains ($a \leq 20$~\micron)
  and a power-law size distribution slope of -3.  The model produces
  good agreement with optical dust polarization data.
  \citeauthor{lasue06} and other investigators using light scattering
  and thermal emission models often describe the porosity of grain
  aggregates with the fractal dimension of the grain.  The density of
  a grain aggregate of homogeneous composition with fractal dimension
  $D$ is described by
  \begin{equation}
    \rho(a) = \rho_0\left(\frac{a}{a_0}\right)^{D-3},
  \end{equation}
  where $\rho_0$ is the bulk density of grain material, and $a_0$ is
  the minimum particle size (typically $a_0 \lesssim 0.1$~\micron).
  \citet{lasue06} found that the scattering properties of their grain
  aggregates were not significantly affected by aggregates with values
  of $D$ ranging from 1.5 to 2.9 [for $\rho_0=2.5$~\gcm:
    $\rho(1~\micron) \approx 0.08-2$~\gcm, $\rho(10~\micron) \approx
    0.003-2$~\gcm].

\item \citet{kolokolova07} also find the scattering properties of
  compact and fluffy grains to be similar, yet the absorption cross
  sections to be significantly different.  They suggest that comets
  with low semi-major axes have dust comae that are comprised of more
  compact grains than comets with large semi-major axes.  Grain
  compactness may be independently revealed by the strength of the
  10~\micron{} silicate emission feature.  Orbit integrations by
  K.~Kinoshita\footnote{\url{http://www9.ocn.ne.jp/\textasciitilde{}comet/}}
  show the semi-major axis of 67P has ranged from 3.5--4.4~AU in the
  past 100~yr, and low-resolution spectrophotometry of the coma of 67P
  by \citet{hanner85} reveal no indication of a silicate feature at
  10~\micron{}.  Together, these facts indicate comet 67P ejects
  compact, rather than fluffy, grains.  \citet{kolokolova07} list the
  vacuum fraction of compact grains to be $\approx0.85$ for equivalent
  volume radii ranging from 0.3--100~\micron{}.

\item The \stardust{} spacecraft returned more than 10,000 dust
  particles in the 1 to 300~\micron{} size range collected from comet
  81P/Wild \citep{brownlee06}.  The spacecraft collected dust grains
  in a porous silica glass ($\rho \leq 0.05$~\gcm) at a relative
  speed of 6.1~\kms{}.  This collection method caused many grains to
  fragment, yet some information on their structure is still retained.
  The collected grains range in densities from $\approx3$~\gcm{} to as
  low as 0.3~\gcm{} \citep{horz06}.
\end{enumerate}

Equation~\ref{eqn:beta} is valid for spherical grains, but greatly
under-dense grain aggregates, such as those discussed above, should
not be expected to scatter light in a manner equivalent to spheres.
\citet{aclr07} computed $\beta$-values of fluffy aggregates using a
combination of Maxwell-Garnet effective medium theory and Mie theory
(for $2\pi a/\lambda \leq 100$) and geometric optics (for $2\pi
a/\lambda > 100$).  The authors found that the $\beta$-values of large
aggregates ($a \gtrsim 1$~mm) are larger than the $\beta$-value of
equivalent volume spheres.  The grain composition plays an important
role for $a \lesssim 1$~mm: for silicate aggregates $\beta$ is less
than or equal to the $\beta$ of equivalent volume spheres; for
amorphous carbon aggregates $\beta$ is greater than or equal to the
$\beta$ of equivalent volume spheres.  Both compositions have
$\beta$-values within an order of magnitude of their equivalent
spheres.  \textit{In situ} evidence from comets 1P/Halley and 81P/Wild
suggest comet grains can be heterogeneous in composition
\citep{fomenkova92, brownlee06, horz06, keller06}.  Heterogeneous
grain models by \citep{kimura06} qualitatively reproduce observed
light scattering properties of dust.  Further work incorporating
heterogeneous compositions into models of light scattering by trail
grains (both aggregate and spheroidal) is needed.

In addition to the differences in $\beta$-values, aggregates should be
ejected from the nucleus at speeds different from equivalent spheres
due to dissimilar grain surface areas per mass.  Altogether,
incorporating a mixture of aggregates and compact grains into a
dynamical model will imbue a range of sizes and velocities on a set of
grains with a given $\beta$-value.  Observational data that could
reveal the structure and composition of trail grains is limited.  We
must necessarily take caution when transforming $\beta$-values to
grain radii and grain masses, but it appears that order of magnitude
estimates are possible.

Equation~\ref{eqn:fnet} describes the net force acting on a dust
particle orbiting the Sun, although the Sun is not the dynamical
center of the solar system.  To appropriately treat a dust particle in
our solar system, our dynamical model also includes the gravitational
accelerations of the planets and, for completeness, the
Poynting-Robertson effect.  We use JPL's planetary ephemeris DE-405
\citep{standish98} to determine the positions of the planets, and the
HORIZONS ephemeris generator \citep{giorgini96} for the comet
positions and velocities.  HORIZONS ephemerides account for
perturbations by the planets and for non-gravitational accelerations
due to comet outgassing.  The net force on each particle is integrated
from ejection to the time of observation using the RADAU-15 integrator
of \citet{everhart85}, which is accurate to a 15th-order series
expansion of the acceleration as a function of time.  The final
particle positions are projected onto the sky for an arbitrary
observer (e.g., Earth or \spitzer{}).

To test the accuracy of the program, we integrated $\beta = 0$
particles released from a comet with a highly eccentric orbit,
28P/Neujmin ($e = 0.78$, $q = 6.91$~AU), and one with a near circular
orbit, 29P/Schwassmann-Wachmann ($e = 0.044$, $q = 5.99$~AU).  A
$\beta = 0$ particle is the large particle limit ($F_{rad} << F_g$)
and represents the nucleus of a comet.  The JPL orbit solutions for
comets Neujmin and 29P/Schwassmann-Wachmann do not include
non-gravitational accelerations, therefore, the positions of $\beta =
0$ particles released with $v_{ej} = 0$ should coincide with the
ephemeris positions of the nuclei.  In 0--20 year integrations the
final positions of $\beta = 0$ particles typically agree within
30--150~km, but increase to $\approx 2000$~km after two perihelion
passages for comet Neujmin.  The accuracy of the program is sufficient
for the following simulations (2000--3000~km corresponds to
$\approx1\arcsec$ in the \spitzer{} image).

\subsection{Model Images}\label{mcmodel}

We employ our dynamical model and a Monte Carlo technique to create
simulated images of comets.  Our simulations consist of 500,000 test
dust particles ejected from comet 67P.  The dust particles are ejected
from the sunward hemisphere of the comet.  The dynamical model's
ejection velocity is
\begin{equation}
v_{ej} = C v_0 \sqrt{\beta / r_h}, \label{eqn:vej}
\end{equation}
where $v_0$ and $v_{ej}$ are in units of \kms, $r_h$ is in units of
AU, and $C$ is an optional scaling factor dependent on insolation.
The $\sqrt{\beta / r_h}$ parameter has been successfully used in other
comet models \citep{lisse98, reach00, ishiguro07} and represents
varying insolation with $r_h$ and the changing effect of gas drag with
$\beta$.  \citet{lisse98} used a variety of other velocity models and
found Eq.~\ref{eqn:vej} to best reproduce the \textit{COBE}
observations of comet comae.

The Monte Carlo model picks particles from a uniform distribution in
time and a logarithmic distribution in $\beta$ ($dn/d\log{\beta}
\propto 1$).  The distributions assure we have useful statistics for
all particle sizes and ages.  For example, if we instead chose a very
steep particle size distribution (PSD) of $a^{-3}$ ($dn/d\beta =
\beta$) where $a$ ranges from 1--$10^4$~\micron{}, the probability of
picking a particle with size $>10^3$~\micron{} is roughly
$1\times10^{-6}$.  The simulated image from a $10^6$ particle
simulation would include only a few $>10^3$~\micron{} sized particles.
The absence of large particles might lead the modeler to conclude that
they do not contribute to the thermal emission from the coma.
Particles of size $10^3$~\micron{} have a larger surface area than
particles of size 1~\micron{}; at the same temperature they emit $\sim
10^6$ times more energy in the thermal infrared.  In addition,
dynamical effects quickly disperse small particles from the vicinity
of the nucleus as the larger particles accumulate.  A simulation that
does not carefully pick particle sizes may misrepresent the relative
contributions of large and small particles.

We also simulate comet dust ejection as constant with heliocentric
distance.  Similar to the PSD of the simulation, this choice ensures
that low levels of activity are appropriately treated.  Our chosen age
distribution (uniform) and PSD ($dn/d\log{\beta} \propto 1$) are
transformed into more appropriate distributions when we generate
simulated images.  The synthetic imager weights each particle so that
the simulated image is representative of a realistic dust production
and PSD (as detailed below).

Simulated observations are necessary to compare the model to the 67P
observations in order to determine if the chosen parameters [$v_{ej}$,
dust production ($Q_d$), grain parameters] describe the images.  After
the simulation is complete, the final particle positions are projected
onto the sky and observed by an imaginary detector, i.e., the total
emission along a line of sight is recorded into an array of pixels.
The simulated image may then be treated and processed as any other
observation, e.g., it may be Gaussian smoothed or unsharp masked.

The projection of the simulation onto the sky gives us the number of
particles in a pixel, $n_g ds$, where $n_g$ is the density of grains
and $ds$ is distance along the line of sight.  The thermal emission
from a collection of particles of uniform temperature, $T$, and size,
$a$, is given by
\begin{equation}
I_\lambda = \int_0^{\infty} \pi a^2 Q_{em}(\lambda,a)\ B_\lambda(T)\ n_g\
ds,
\label{eqn:simuemiss}
\end{equation}
where $B_\lambda(T)$ is the Planck function.  The emission
coefficient, $Q_{em}$, is roughly $2\pi a / \lambda$ for $a \lesssim
\lambda / 2\pi$, and of order unity for $a \gtrsim \lambda / 2\pi$.
The dependence of $Q_{em}$ on mineralogy and structure is beyond the
scope of this treatment.  Combining all constant factors and
transforming our integral into a summation, we now have
\begin{equation}
I_\lambda \propto \sum_{i = 1}^N a_i^2
	\begin{cases} 2 \pi a_i / \lambda &\text{for } a_i \lesssim
                        \lambda / 2\pi\\
                      1             &\text{for } a_i \gtrsim
                        \lambda / 2\pi,
	\end{cases} \label{eqn:thermeq}
\end{equation}
where $N$ is the number of particles along the line of sight.

Real comets do not eject particles with $dn/d\log{\beta} \propto 1$ or
$dn/dt \propto 1$.  We remove the $dn/d\log{\beta}$ bias and weight
each particle by the overall dust production trend and an ejected PSD.
The dust production will commonly be a power-law of the form, $Q_d
\propto r_h^{-k}$ \citep[e.g.,][]{ahearn95}, where $r_h$ is the
heliocentric distance at particle release and $k$ is the logarithmic
slope of the heliocentric dust production (nominally, -4).  For now,
we leave the PSD as an arbitrary function, $n(\beta)|_{PSD}$.  Each
test particle, $i$, now represents a collection of particles, $n_i$:
\begin{equation}
n_i \propto \frac{
  \left. n(\beta_i) \right|_{PSD}}{
  \left. n(\beta_i) \right|_{sim}}
= \frac{
  \left. \int \frac{dn}{d\beta} d\beta \right|_{PSD,i}}{
  \left. \int \frac{dn}{d\beta} d\beta \right|_{sim,i}}
\approx \frac{
  \left. \frac{dn}{d\beta} \right|_{PSD,i}}{
  \left. \frac{dn}{d\beta} \right|_{sim,i}}
= \beta_i \left. \frac{dn}{d\beta} \right|_{PSD,i},
\label{eqn:psdeq}
\end{equation}
where $n(\beta)|_{sim}$ is the PSD of the simulation (here,
$dn/d\log{\beta}$).  Combining Eq.~\ref{eqn:thermeq} and
\ref{eqn:psdeq} yields the intensity of thermal emission falling onto
a pixel,
\begin{eqnarray}
I_\lambda & \sim & \sum_{i = 1}^N n_i\
           Q_{d,i}\ Q_{em,i}(\lambda,a_i)\ A_i \nonumber \\
  & \sim & \sum_{i = 1}^N r_{h,i}^{-k} \beta_i \left. \frac{dn}{d\beta} \right|_{PSD,i}
	   \begin{cases}
	     2 \pi / (\beta_i \lambda)
             & \text{ for } a_i \lesssim \lambda/2\pi \\
  	     1
	     & \text{ for } a_i \gtrsim \lambda/2\pi
	   \end{cases}.
\label{eqn:dynaI}
\end{eqnarray}
In \S\ref{sec:psd} we consider various PSDs for the comet.  At
present, we apply an interstellar medium PSD, $dn/da \sim a^{-3.5}$
\citep{mathis77}, and use $\rho=1$~g~cm$^{-3}$ to transform between
$a$ and $\beta$.

\subsection{Ejected Dust Velocity}\label{sec:modelvej}
The forward emission along the trail of 67P can be explained by one of
two possibilities.  The emission consists of: 1) very old trail
particles that will soon be lapped by the nucleus, or 2) grains
ejected at some appreciable velocity from the nucleus that fall ahead
of the comet.  \textit{IRAS} observations of 67P show a forward trail
extending $0.42 \times 10^6$~km \citep{sykes92} and \spitzer{}
(Fig.~\ref{fig:profile}) shows forward emission $1.2 \times 10^6$~km
in length.  If very old trail particles fill the entire orbit and are
being lapped by the comet nucleus, then a longer extension, continuing
around the orbit, should have been observed by both \textit{IRAS} and
\spitzer{}.  Moreover, filling an entire cometary orbit with trail
particles could take hundreds of years \citep{reach00}.  Over this
time the planets will have many opportunities to separate the comet
nucleus and trail particles through minor and major gravitational
perturbations.  The separation of trail and nucleus has been observed
through meteoroid stream studies of comet 8P/Tuttle and
55P/Tempel-Tuttle \citep{jenniskens02, jenniskens00} and modeled in a
study of 67P's meteoroid stream \citep{vaubaillon04}.  We also note that
comet 67P's last encounter with Jupiter occurred in 1959 (at an
encounter distance of 0.05~AU).  We conclude that it is highly
unlikely that very old trail particles are being lapped by the
nucleus.

To better understand the effect of ejection velocity on ejected dust
grains, we use our dynamical model to simulate observations of comet
67P at the same observation time, viewing geometry, and wavelength as
our 24~\micron{} image.  Particle ages range from 0--4000 days and the
$\beta$ parameters range from $10^{-5}$--$10^{-1}$ ($a \approx 6$ to
$6\times10^{4}$~\invrho~\micron{}).  A maximum age of 4000~days
corresponds to 1.7 orbital periods (March 1993, $r_h=5.65$
pre-perihelion).  Including the previous orbit allows us to place
limits on trail ejection velocity and dust production, if we consider
that one orbital period is the minimum age of a particle to be
considered a trail particle.  We test three approaches to represent
the ejection velocity ($v_{ej}$) and dust production ($Q_d$) from the
surface of the nucleus (via parameter $C$ in Eq.~\ref{eqn:vej}): 1)
$Q_d$ and $v_{ej}$ are uniform across the sunward hemisphere, 2) $Q_d$
and $v_{ej}$ are proportional to insolation, $\cos{z_\sun}$, where
$z_\sun$ is the Sun-zenith angle, and 3) $Q_d$ and $v_{ej}$ are
proportional to the surface temperature of a slowly rotating nucleus,
$\cos^{1/4}{z_\sun}$.  An example of each approach is presented in
Fig.~\ref{fig:sims-v0.5} for $v_0 = 0.5$~\kms{} (Eq.~\ref{eqn:vej})
and $dn/da \sim a^{-3.5}$.

The uniform model (method~1) produces an abundance of dust with an
ejection velocity perpendicular to the orbital plane and causes the
observed dust morphology to be dispersed perpendicular to the
projected orbit.  A lower $v_0$ can be used to suppress this effect,
but lower velocities do not eject dust in the forward direction to the
extent observed by \spitzer{} in 4000~day simulations
(Fig.~\ref{fig:mips}).  The introduction of a cosine term into $Q_d$
and $v_{ej}$ (methods~2 and~3) reduces the extent of the dust emission
in the perpendicular direction, by 1) reducing $Q_d(z_\sun=90\degr)$
to zero, and 2) reducing $v_{ej}(z_\sun\gtrapprox30\degr)$ to lower
velocities.  Both $(Q_d, v_{ej}) \propto \cos{z_\sun}$ and $\propto
\cos^{1/4}{z_\sun}$ exhibit a smaller coma width in
Fig.~\ref{fig:sims-v0.5} and are preferred over the wide comae
produced in the uniform production approach.  The $\cos{z_\sun}$ and
$\cos^{1/4}{z_\sun}$ approaches appear equally viable, but for
simplicity we elect to adopt $\cos{z_\sun}$ ($Q_d$ and $v_{ej}$
proportional to insolation) for the duration of the analysis.

We calculated our $C=\cos{z_\sun}$ model with a range of ejection
velocities, $v_0 = 0.1$, 0.2, 0.3, 0.5, 1.0, and 1.5~\kms{}
(Fig.~\ref{fig:sims-vej}).  Varying $v_0$ affects the extent and
profile of the forward emission and the width and intensity of the
backward emission.  The 0.5~\kms{} model best matches the 24~\micron{}
forward emission, backward emission, and the small inner-coma,
although the inner-coma is not exactly reproduced.  The lowest
ejection velocity did not produce the forward emission in 4000~days
and larger velocities created broad coma and backward emission.

\subsection{The Particle Size Distribution}\label{sec:psd}

We simulated images for a variety of power-law PSDs
(Fig.~\ref{fig:sims-psd}) to examine the effect of the comet's ejected
PSD on the dust morphology.  A single power-law may not represent the
comet's ejected PSD, but it suffices as a first approximation.  We
chose the PSD to be $dn/da \propto a^k$, where $k = -5.0$ to $-0.5$ in
steps of 0.5 ($dn/d\beta \sim \beta^{-(k+2)}$).  Note that the chosen
PSDs represent the \textit{ejected} dust and not the \textit{observed}
dust.  The observed PSD is a result of the dynamical separation of
dust after ejection, i.e., the observed PSD has an enhanced large
particle portion as small particles are quickly removed from the coma
and large particles accumulate over time.  The ejected PSD that best
matches the observed 24~\micron{} image (Fig.~\ref{fig:mips}) is
$dn/da \sim a^{-3.5}$.  The $a^{-3.5}$ model reproduces the observed:
1) forward emission, 2) backward emission, which increases in width
with distance behind the nucleus, and 3) approximately flat
perpendicular surface brightness profile at $\approx500$\arcsec{} from
the nucleus.  \citet{fulle04} found a time-dependent power-law index
best-fit the 67P photometry using the inverse tail approach
\citep{fulle89}, but their derived time-average of $k = -3.4$ is very
close to our derived $k \approx -3.5$.  Other comets have $k = -3.0$
to --4.1 as determined with the inverse tail approach
\citep{fulle04a}.  A summary of our best model's parameters is
presented in Table~\ref{table:best-model}.

We also simulated the coma of 67P at the epoch of our Palomar
observation.  Figure~\ref{fig:sim-pal} presents the simulation with
the same viewing geometry as the 0.6~\micron{} image
(Fig.~\ref{fig:lfcimage}).  The model parameters were the same as the
best 24~\micron{} simulation (Table~\ref{table:best-model}), but the
image is simulated at a wavelength of 0.6~\micron, instead of
24~\micron{} (since we are assuming isothermal grains and limit
$a>0.5$~\micron{}, Eq.~\ref{eqn:thermeq} also serves as a rough
approximation for scattered light).  The resulting simulation shows a
coma, and two thin dust structures: one structure is located along the
comet's projected orbit, and the other is located to the north of the
projected orbit.  The 0.6~\micron{} image does not does not show the
emission along the projected orbit, but does exhibit the thin line of
emission to the north.  The emission along the projected orbit is
comprised of particles with a lower $\beta$-value than the thin line
of emission just to the north of the orbit.  Slightly decreasing the
PSD slope from $k = -3.5$ could make the lower $\beta$ structure
undetectable in the 0.6~\micron{} images.  Alternatively, the PSD
could be better described by a multi-component power-law, such as
those measured \textit{in situ} at comets 1P/Halley and 81P/Wild
\citep{mcdonnell87, green04}.  Modifying the PSD would have important
effects on the morphology at the 24~\micron{} image epoch, but varying
the dust production or ejection velocities may recover the simulated
emission.  In a third option, we could simply vary the dust production
at the epoch when the northern dust feature is ejected.  As is, the
parameters of Table~\ref{table:best-model} adequately represents both
the 0.6 and 24~\micron{} images to within their quality.

\subsection{The Nature of the Simulated and Observed
Emissions}\label{sec:dustnature}

We compared the photometry profiles derived from the 24~\micron{}
image to profiles derived from the best simulated 24~\micron{} image
(Fig.~\ref{fig:sim-profile}) and found that the 24~\micron{}
profile parameters listed in Table~\ref{table:profilefit}
approximate the simulated image profile.  This gives confidence in the
simulation and that one may draw qualitative conclusions on the nature
of the observed emission.

\citet{fulle04} identified comet 67P to have a persistent neck-line
tail structure beginning $\approx60$~days after the August 2002
perihelion passage.  The neck-line structure is a thin enhancement of
the projected dust surface density in the tail created when dust
ejected perpendicular to the plane of the orbit crosses the plane
again at the node 180\degr{} away (i.e., a true anomaly, $f$,
difference of 180\degr) from its point of emission.  The enhancement
can only be detected when the observer is located fairly close to the
orbital plane of the comet.  Comet 67P's orbit inclination is
7\degr{}, which is very favorable for neck-line observations.  The age
of dust ejected at $\Delta f = 180$\degr{} from the \spitzer{}
observation (neck-line dust) is 590~days.  This age corresponds to
dust ejected 36~days before the August 2003 perihelion, during
significant coma activity (\afrho{} $\approx1000$~cm).  We decomposed
our best 24~\micron{} simulation into three age bins
(Fig.~\ref{fig:bestmodel}): 1) dust ejected before the May 2000
aphelion, 2) dust ejected at 16--56~days before the August 2003
perihelion, and 3) the remaining dust.  Bin~1 corresponds to trail
particles, is centered on 67P's projected orbit, has a mean
$\beta$-value $\overline{\beta} = 1.03\times10^{-4}$ and a mean radius
$\overline{a} = 8470$~\micron.  Bin~2 corresponds to the neck-line
tail structure located to the south of the projected orbit and has
similar sized particles as the trail: $\overline{\beta} =
1.03\times10^{-4}$, $\overline{a} = 7790$~\micron.  A comparison of
Figs.~\ref{fig:mips}, \ref{fig:perp-profile}, and \ref{fig:bestmodel}
reveals that the forward emission in the \spitzer{}/MIPS image is the
dust trail, and the backward emission has contributions from both the
dust trail and neck-line.  We also examined the 0.6~\micron{} image
and simulation (Fig.~\ref{fig:sim-pal}) and identify the thin line of
emission to the north of the projected orbit as a neck-line.  The LFC
image was not sensitive enough to detect the dust trail.

\subsection{\textit{Rosetta}'s Dust Impact Hazard}\label{sec:impact}
During encounters with comets, spacecraft are vulnerable to dust
impacts.  The \rosetta{} spacecraft is designed to follow and approach
comet 67P and subsequently orbit the comet nucleus at distances of 5
to 25 nucleus radii.  Orbit insertion missions have a lower impact
hazard than typical flyby missions due to the lower encounter
velocities (m~s\inv{} versus km~s\inv{}).  The largest particles ($a >
10^3$~\micron) pose the greatest threat to spacecraft health, but
typical comet PSDs suggest these particles are fewest in number.  To
predict the large particle impact hazard to \rosetta{}, we estimate
the surface brightness of 67P's dust trail near the comet nucleus and
then calculate a trail grain density.  The 24~\micron{} image at the
nucleus has a surface brightness of $0.61\pm0.10$~\mjysr{}
(Fig.~\ref{fig:pointsource}).  \rosetta{} will encounter 67P while the
comet is approaching the Sun at $r_h = 4-5$~AU.  At orbit insertion,
the nucleus will be surrounded by persistent trail dust, but, since
little or no dust production is expected near aphelion, the transient
neck-line and coma dust will have dissipated.  According to our best
model, the trail comprises 12\% of the total dust emission near the
nucleus, or $0.07\pm0.01$~\mjysr{} in the 24~\micron{} image.  The
error quoted for the trail component arises from the formal
measurement uncertainty of the dust surface brightness.  The error due
to model uncertainties is difficult to estimate.  We find a comparable
surface brightness in the forward trail ($\approx0.1$~\mjysr), which
suggests that we have the correct order of magnitude.  A trail surface
brightness $\sim0.1$~\mjysr{} serves as a useful estimate near the
nucleus.  This value corresponds to a grain number density of
$\sim10^{-11}$~m$^{-3}$.  We have assumed the trail consists entirely
of 1~mm sized grains, the dust temperature is 300$\sqrt{r_h}$~K,
derived from \iras{} color-temperatures of comet dust trails
\citep{sykes92}, and the near-nucleus volume is
$22,000\times22,000\times60,000$~km.  The volume corresponds to the
spatial resolution of the MIPS instrument and the trail width measured
in \S\ref{sec:profiles}.  Using a spacecraft cross-section of
4~m$^{2}$ \citep{esa07} and a path length of half the trail width
(30,000~km), we calculate that \rosetta{} has a $\sim0.1$\% chance of
encountering a 1~mm sized trail particle during orbit insertion.  The
low probability is expected.  During comet flyby missions, spacecraft
typically encounter a few $0.1-1$~mg ($a\sim500$~\micron) dust grains
\citep{green04, mcdonnell87, mcdonnell93}, but these encounters occur
during vigorous coma activity, which enhances the probability of large
particle impacts.

\section{Conclusions}\label{sec:conclusions}

Comet 67P/Churyumov-Gerasimenko's dust trail was detected in
\spitzer{}/MIPS images when the comet was at 4.5~AU post-perihelion.
By comparison of a 24~\micron{} image to dynamic models, we estimate
that the trail particles are large ($\gtrsim 100$~\micron) and old (at
least 1 orbit or 6.6 years).  They are ejected at relative velocities
of $\approx 0.5 \sqrt{\beta/r_h} \cos{z_\sun}$~\kms{}, or of order
1~m~s$^{-1}$ at perihelion, although we note that the ejection
velocity may be lower, if older, and slower, particles are included in
the dust trail.  The simulations show two distinct features.  One
feature is centered on the projected orbit of the comet and extends in
the forward and trailing directions.  Given the low average $\beta$
($10^{-4}$) and large age (up to the limits of the simulation) we
identify the feature to be the comet's dust trail.  The other feature
is located at a few degrees south of the trail in position angle (at
the \spitzer{} epoch/viewing geometry) and extends behind the comet.
This feature is the ``neck-line'' structure and is also comprised of
low $\beta$ ($10^{-4}$) grains, but the orientation and age (590~days
at the \spitzer{} epoch) preclude it from being identified as a dust
trail.  The morphology of the \spitzer{} (24~\micron) and Palomar
(0.6~\micron) images agree with our dynamical model.  In particular,
the forward trail is observed in the 24~\micron{} image, and the
neck-line is observed in both images.  The dust trail is found behind
the comet, but is overwhelmed by emission from the neck-line dust.
The dust trail is below the detection limit of our optical image.

Comet dust particle size distributions favor small particles
($a\lesssim10$~\micron) by number, although larger particles may still
dominate the ejected dust mass.  Consequently, small particles are
more readily detected in optical images and typical 10~\micron{}
spectra of comets.  Observations of large particles, dynamically
separated from comet comae and tails, allow for a detailed study of
large dust grains ejected from comet nuclei.  In the study of comet
67P's dust trail, we find that the ejected dust grain size
distribution is approximately a power-law (slope $-3.5$) for particle
sizes from 1 to $10^4$~\micron{}, or more.  Thermal infrared spectra
\citep{harker05, harker02, lisse06} and dust collectors on-board comet
flyby spacecraft \citep{mcdonnell87, green04} also measure grain size
distributions with power-law slopes in the --3 to --4 range, with some
variation between models and particle size bins.  Such power-law
slopes have important implications on the total mass ejected from
comet 67P.  With our PSD derived for 67P, the mass of particles
studied by thermal models of mid-IR spectra ($a=0.1$ to
$\approx20$~\micron) comprise approximately 10\% of the total ejected
dust, when particles up to $10^3$~\micron{} are considered.  Using our
dynamical model, we estimate the surface brightness of the dust trail
near the nucleus and find \rosetta's probability of impacting a dust
trail grain to be $\sim10^{-3}$, assuming the grains are all
$10^3$~\micron{} in radius.  The low probability of impact coupled
with the low relative velocities required for orbit insertion suggests
that 67P's dust trail poses a minor threat to the \rosetta{}
spacecraft.

\section*{Acknowledgements}
The authors thank the reviewers, C.~M. Lisse and an anonymous
reviewer, for comments that greatly improved the manuscript.  The
authors also thank C.~E. Woodward for reviewing the paper.  MSK thanks
R.~Gal for sharing Palomar Large Format Camera reduction techniques.
MSK acknowledges partial support from the National Science Foundation
grant AST-037446 and the \textit{Spitzer} Science Center's Visiting
Graduate Student Fellowship Program.  Support for this work was
provided by NASA through Contract Number 1263741 issued by JPL/Caltech
to the University of Minnesota.  This work is based in part on
observations made with the \textit{Spitzer Space Telescope}, which is
operated by the Jet Propulsion Laboratory, California Institute of
Technology under NASA contract 1407.  This work is also based in part
on observations obtained at the Hale Telescope, Palomar Observatory,
as part of a collaborative agreement between the California Institute
of Technology, its divisions Caltech Optical Observatories and the Jet
Propulsion Laboratory (operated for NASA), and Cornell University.


\clearpage


\begin{deluxetable}{llccccc}
\tablecolumns{7}
\tablecaption{Comet 67P/Churyumov-Gerasimenko trail observation
details.\label{table:obsgeom}}
\tablewidth{0pt}
\tablehead{
  &
  &
  &
  &
  &
  \colhead{Phase} &
  \colhead{Projection} \\
  &
  \colhead{Date /} &
  \colhead{$T-T_p$\tablenotemark{a}} &
  \colhead{$r_h$} &
  \colhead{$\Delta$\tablenotemark{b}} &
  \colhead{Angle} &
  \colhead{Angle\tablenotemark{c}} \\
  \colhead{Observatory} &
  \colhead{Time (UT)} &
  \colhead{(days)} &
  \colhead{(AU)} &
  \colhead{(AU)} &
  \colhead{(degrees)} &
  \colhead{(degrees)}
}

\startdata
\iras{} & 1983 May 05 13:00 & 174.4 & 2.28 & 2.09 & 26 & 8 \\
Palomar & 2003 Jun 26 04:36\tablenotemark{d} & 312.0 & 3.23 & 3.35 & 18 & 20\\
\spitzer{} & 2004 Feb 23 03:44 & 553.9 & 4.47 & 4.06 & 12 & 48 \\
\enddata

\tablenotetext{a}{Time since the closest perihelion.}

\tablenotetext{b}{Observer-comet distance.}

\tablenotetext{c}{Angle between the comet's heliocentric velocity
vector and the image plane.}

\tablenotetext{d}{Average of three observations: June 25.1840, 26.2049,
27.1861.}

\end{deluxetable}

\begin{deluxetable}{lr@{ $\pm$ }lr@{ $\pm$ }lr@{ $\pm$ }ll}
\tablecolumns{8}
\tablecaption{Best-fit, de-projected dust profiles of the comet
67P/Churyumov-Gerasimenko \spitzer{}/MIPS observation
(Fig.~\ref{fig:profile}).\tablenotemark{a}
\label{table:profilefit}}
\tablewidth{0pt}
\tablehead{
  \colhead{Emission} &
  \multicolumn{2}{c}{Slope, $k$} &
  \multicolumn{2}{c}{Scale, $S_\nu$(\mjysr)} &
  \multicolumn{2}{c}{Scale, $\tau$} &
  \colhead{$\chi^2_\nu$}
}

\startdata
0.6~\micron{} backward & -1.03 & 0.02 & 2200 & 400 & 130 & 20 & 4.6 \\
24~\micron{} backward & -0.50 & 0.05 & 350 & 20 & 1.23 & 0.08 & 20.2 \\
24~\micron{} forward & -0.85 & 0.04 & 7000 & 3400 & 25 & 12 & 0.7 \\
\enddata

\tablenotetext{a}{Profile $=Cd^k$, where $d$ is in units
of km, and $C$ is the scale factor.}

\end{deluxetable}

\begin{deluxetable}{ll}
\tablecolumns{2}
\tablecaption{Simulation (Fig.~\ref{fig:bestmodel}) parameters that
best reproduce the \spitzer/MIPS image of comet
67P/Churyumov-Gerasimenko (Fig.~\ref{fig:mips}).
\label{table:best-model}}
\tablewidth{0pt}
\tablehead{
  \colhead{Parameter} &
  \colhead{Value\tablenotemark{a}}
}

\startdata
Dust production, $Q_d$ & $\sim r_h^{-5.8} \cos{z_\sun}$ \\
Particle size distribution, $dn/da$ & $\sim a^{-3.5}$ \\
Ejection velocity, $v_{ej}$ & $0.5 \sqrt{\beta/r_h} \cos{z_\sun}$ (\kms) \\
\enddata

\tablenotetext{a}{The parameter $z_\sun$ is the Sun-zenith angle, dust
production is zero on the night hemisphere ($z_\sun > 90$\degr).}

\end{deluxetable}

\clearpage


\begin{figure}
\caption{Observer-comet 67P/Churyumov-Gerasimenko viewing geometries
for the \textit{IRAS}, Palomar, and \spitzer{} epochs, as seen from
the north ecliptic pole.  The observer positions (\textit{filled} and
\textit{open circles}) are labeled with the telescope names and the
respective comet positions (\textit{triangles}) are labeled with the
telescope names in parentheses.  The orbits of Earth, Jupiter, and
67P/Churyumov-Gerasimenko are also shown. \label{fig:orbit}}
\end{figure}

\begin{figure}
\caption{\spitzer/MIPS 24~\micron{} image of comet
67P/Churyumov-Gerasimenko (\textit{top}) and smoothed
\textit{contours} (\textit{bottom}).  The observation was taken on
2004 February 23 at 03:44 UT.  The neck-line tail structure and dust
trail (identified in \S\ref{sec:dustnature}) are indicated by the
\textit{arrows}.  The first contour is $1.5\sigma$ above the
background and the second contour is selected to enhance the
neck-line.  \textit{Short arrows} denote the projected heliocentric
velocity ($v$) and sunward direction ($\sun$). \label{fig:mips}}
\end{figure}

\begin{figure}
\caption{Mt.\mbox{} Palomar 5~m/Large Format Camera $r^\prime$
(0.6~\micron{}) image of comet 67P/Churyumov-Gerasimenko taken on
2003 June 26 at 04:36 UT.  \textit{Arrows} mark the projected
heliocentric velocity ($v$) and sunward direction
($\sun$). \label{fig:lfcimage}}
\end{figure}

\begin{figure}
\caption{Cuts along the orbit of comet 67P/Churyumov-Gerasimenko
centered on the nucleus at 24~\micron{} for both the original image
(\textit{solid line}) and the point source subtracted image
(\textit{dashed line}).  Each line represents a 3~pixel width average.
The total point source flux used was 2.95~mJy, the mid-point of our
upper- and lower-limits on the flux from the nucleus.  \textit{Inset:}
A $40\times40$ pixel close-up of the nucleus.
\label{fig:pointsource}}
\end{figure}

\begin{figure}
\caption{Peak surface brightness of comet 67P/Churyumov-Gerasimenko as
a function of distance along the de-projected dust emission for the
0.6~\micron{} (\textit{left}) and the 24~\micron{} (\textit{right})
images (see \S\ref{sec:profiles} for details).  The plotted
profiles are described in Table~\ref{table:profilefit}.  All data
points have a signal-to-noise ratio greater than 3 with the exception
of the three data points on the far left of the 0.6~\micron{}
figure. \label{fig:profile}}
\end{figure}

\begin{figure}
\caption{Cuts perpendicular to the projected orbit of comet
67P/Churyumov-Gerasimenko in the point source subtracted \spitzer/MIPS
image.  Labels denote distance from the nucleus, positive values are
in the trailing direction.  The emission ahead of the nucleus is
centered on the projected orbit, meanwhile the emission behind the
nucleus widens as the neck-line structure separates from the orbit and
trail in projection
(\S\ref{sec:dustnature}). \label{fig:perp-profile}}
\end{figure}

\begin{figure}
\caption{Simulated 24~\micron{} images of comet
67P/Churyumov-Gerasimenko with the same viewing geometry as
Fig.~\ref{fig:mips}.  The dust production ($Q_d$) and the magnitude
and direction of $v_{ej}$ is varied between the three panels, but
$v_0$ (Eq.~\ref{eqn:vej}) is held constant at 0.5~\kms: \textit{top
panel}, uniform $Q_d$ and $v_{ej}$ across the sunlit hemisphere;
\textit{center panel}, $Q_d$ and $v_{ej}$ are proportional to
$\cos{z_\sun}$, where $z_\sun$ is the Sun-zenith angle; and
\textit{bottom panel}, $Q_d$ and $v_{ej}$ are proportional to the
surface temperature of a spherical nucleus $(\cos^{1/4}{z_\sun})$.
The images are plot with the same logarithmic scale and have been
Gaussian smoothed (3~pixel Gaussian full-width at half-maximum at
2.5~arcsec~pixel\inv).  The dust production varies as $r^{-5.8}$ and
the particle size distribution is $dn/da \propto a^{-3.5}$.  See
\S\ref{sec:modelvej} for a discussion.
\label{fig:sims-v0.5}}
\end{figure}

\begin{figure}
\caption{Same as Fig.~\ref{fig:sims-v0.5} but only the
$\cos{z_\sun}$ model is shown and $v_0$ is varied (Eq.~\ref{eqn:vej}).
See \S\ref{sec:modelvej} for a discussion. \label{fig:sims-vej}}
\end{figure}

\begin{figure}
\caption{Same as Fig.~\ref{fig:sims-vej} but only the $v_0 =
0.5$~\kms{} model is shown and the ejected particle size distribution
is varied ($dn/da \sim a^k$).  See \S\ref{sec:psd} for a
discussion. \label{fig:sims-psd}}
\end{figure}

\begin{figure}
\caption{Simulated image (\textit{top}) for the same viewing geometry
as the 0.6~\micron{} Palomar/LFC image (\textit{bottom}) using the
parameters of the best 24~\micron{} model
(Table~\ref{table:best-model}).  The image is simulated for a
wavelength of 0.6~\micron{} at 0.7~arcsec~pixel\inv{} (see
\S\ref{sec:psd}).  Both images have been convolved with a 12\arcsec{}
full-width at half-maximum Gaussian and are plot on a linear scale.
The projected orbit of comet 67P/Churyumov-Gerasimenko is shown as a
\textit{solid line}.  The trail (along the orbit) and neck-line (north
of the orbit) are both visible in this simulated image but only the
neck-line appears to be visible in the LFC image
(\S\ref{sec:dustnature}). \label{fig:sim-pal}}
\end{figure}

\begin{figure}
\caption{Surface brightness profile of the best simulated 24~\micron{}
image (Fig.~\ref{fig:bestmodel}).  The \textit{dotted lines} are
scaled versions of the best-fit forward and backward 24~\micron{}
profiles (Table~\ref{table:profilefit}).  The scaling factors used
were 0.021 for the backward emission, and 0.026 for the forward
emission. \label{fig:sim-profile}}
\end{figure}

\begin{figure}
\caption{Same as Fig.~\ref{fig:sims-vej} but only for $v_0 =
0.5$~\kms{} and the emission has been enhanced by a factor of 2 to
increase the image contrast.  The projected orbit is plot as a
\textit{solid line}.  The \textit{top panel} shows the thermal
emission from particles released during the 1996 perihelion (particle
age $> 1750$~days, or 4.8~years), i.e., trail particles.  The
\textit{center panel} shows the thermal emission from particles
released at a true anomaly 180\degr{} prior to the 24~\micron{}
observation ($570 < \mbox{age} < 610$~days), i.e., particles that form
a neck-line tail structure.  Both of the trail and neck-line
components are visible in the 24~\micron{} image,
Fig.~\ref{fig:mips}.  The \textit{bottom panel} shows the remaining
dust after removing the neck-line and trail
components. \label{fig:bestmodel}}
\end{figure}

\clearpage
\setcounter{figure}{0}

\begin{figure}
\plotone{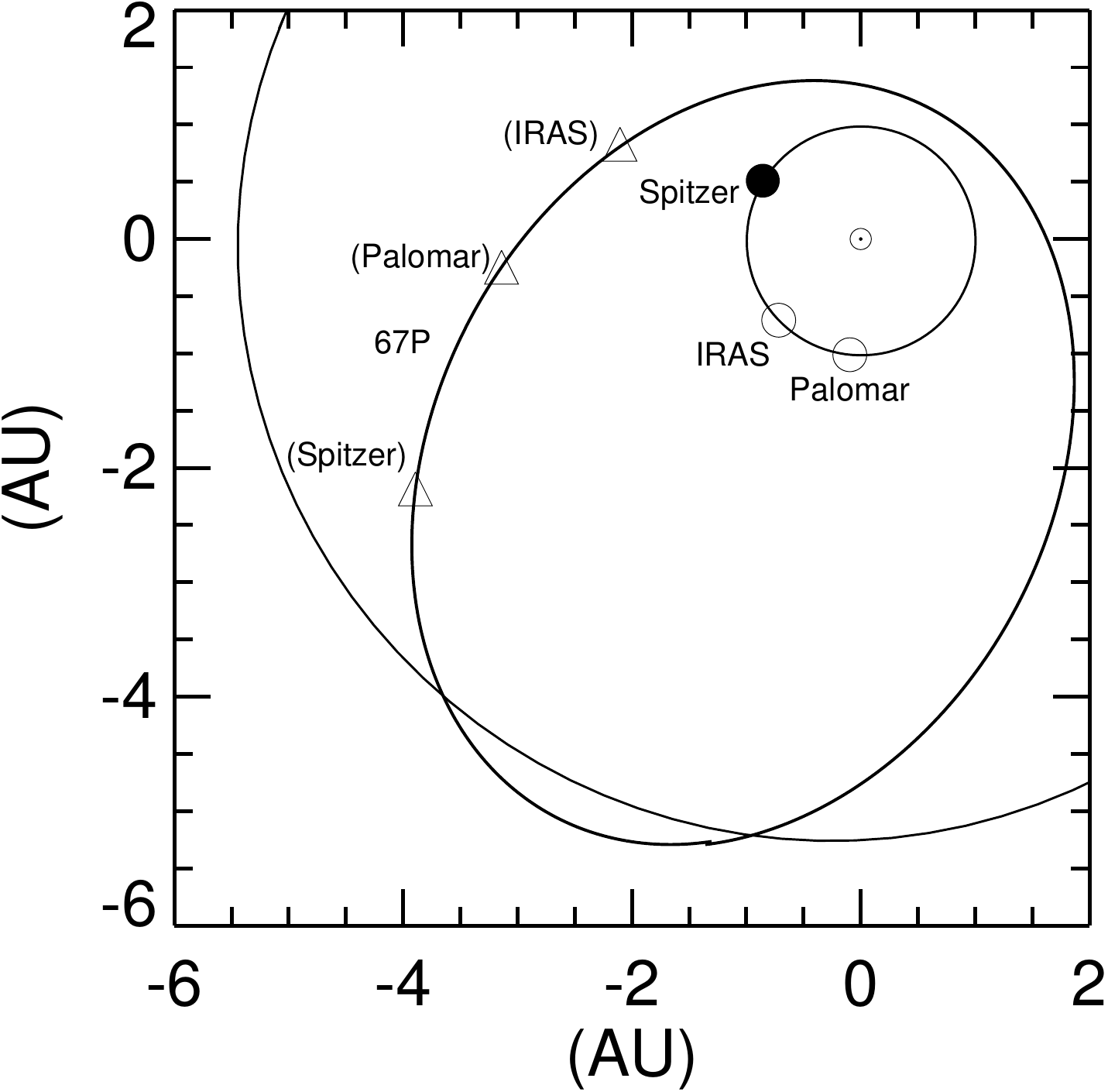}
\caption{Kelley \textit{et al.}, The Dust Trail of Comet 67P/Churyumov-Gerasimenko}
\end{figure}

\begin{figure}
\plotone{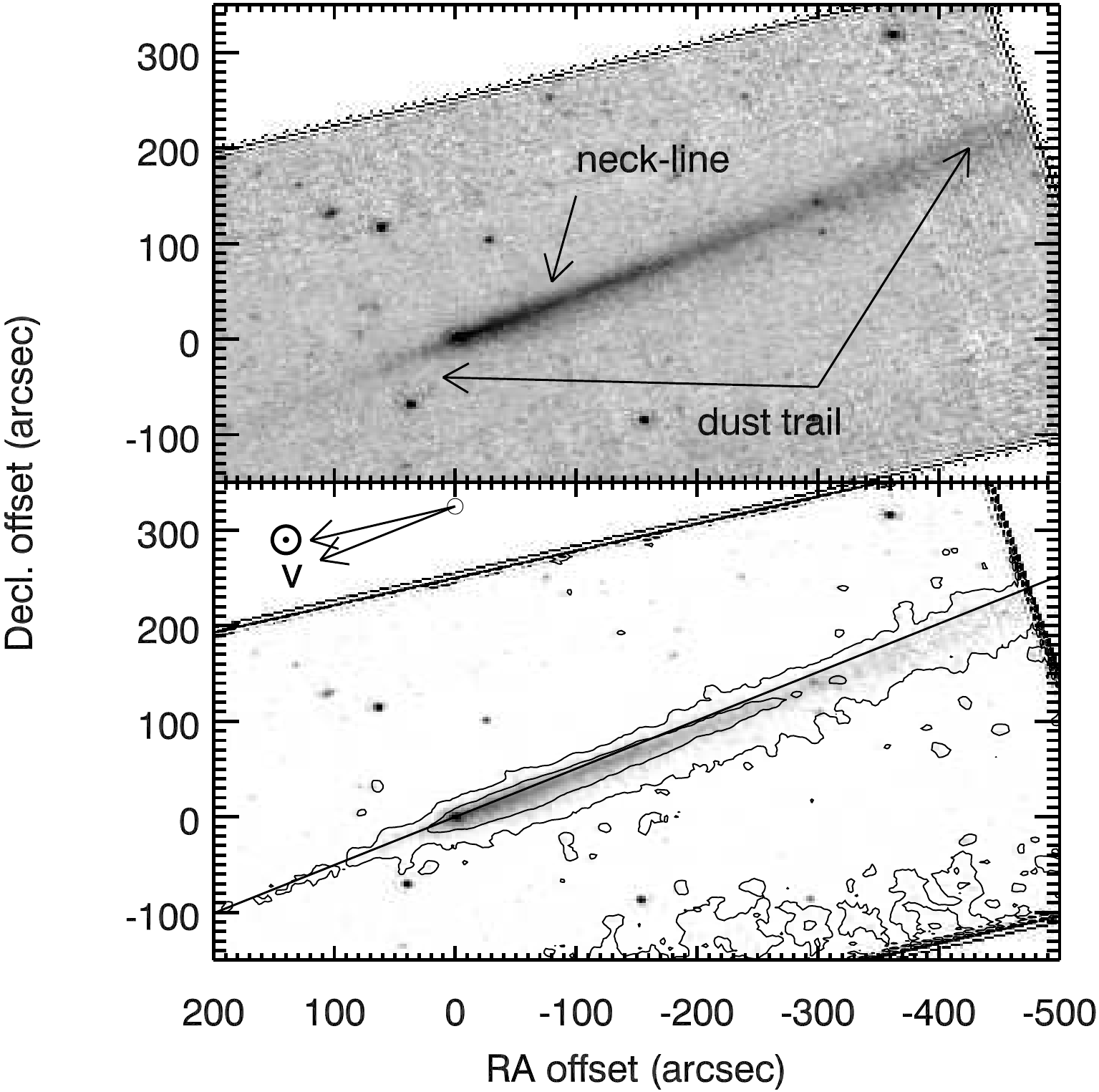}
\caption{Kelley \textit{et al.}, The Dust Trail of Comet 67P/Churyumov-Gerasimenko}
\end{figure}

\begin{figure}
\plotone{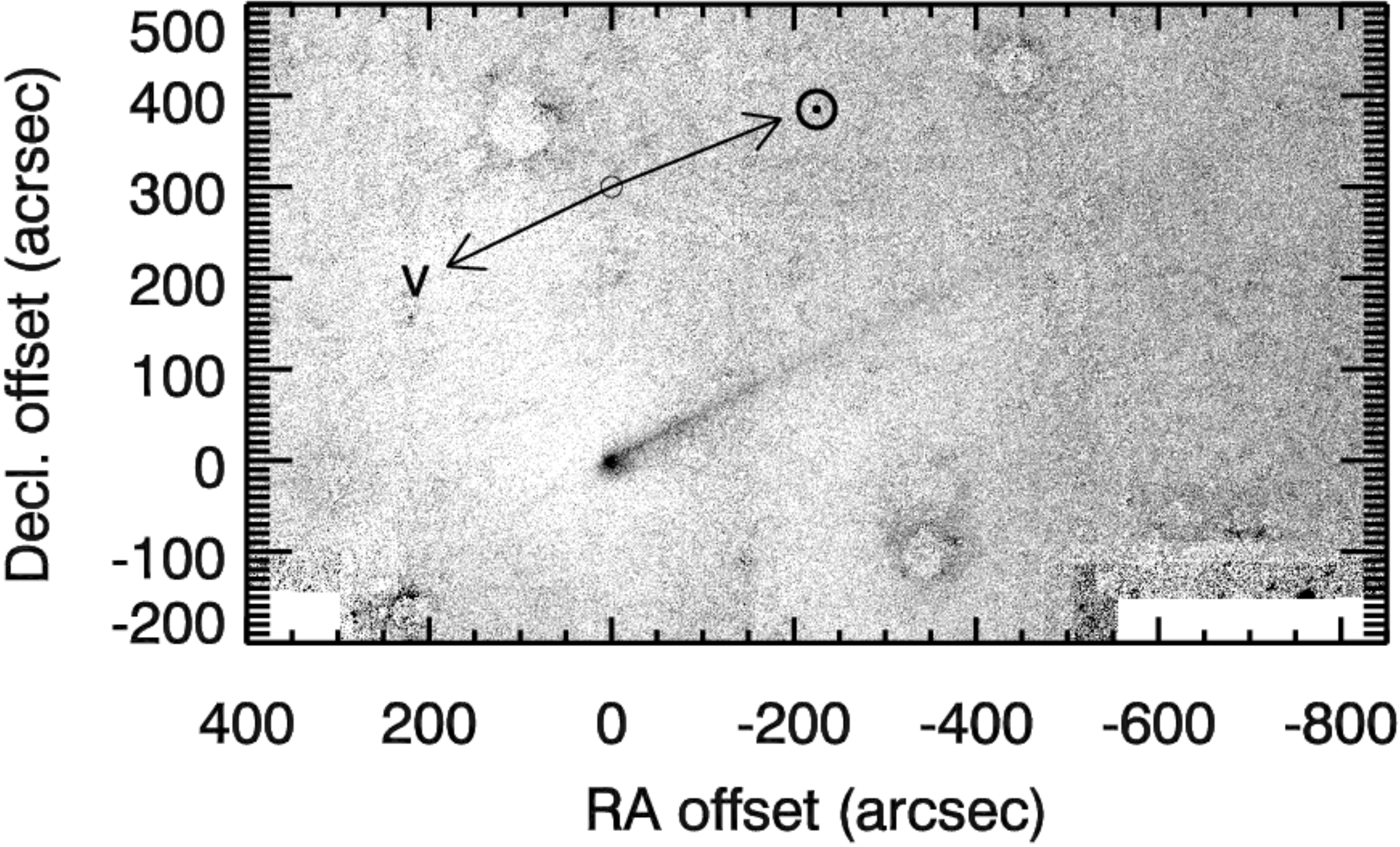}
\caption{Kelley \textit{et al.}, The Dust Trail of Comet 67P/Churyumov-Gerasimenko}
\end{figure}

\begin{figure}
\plotone{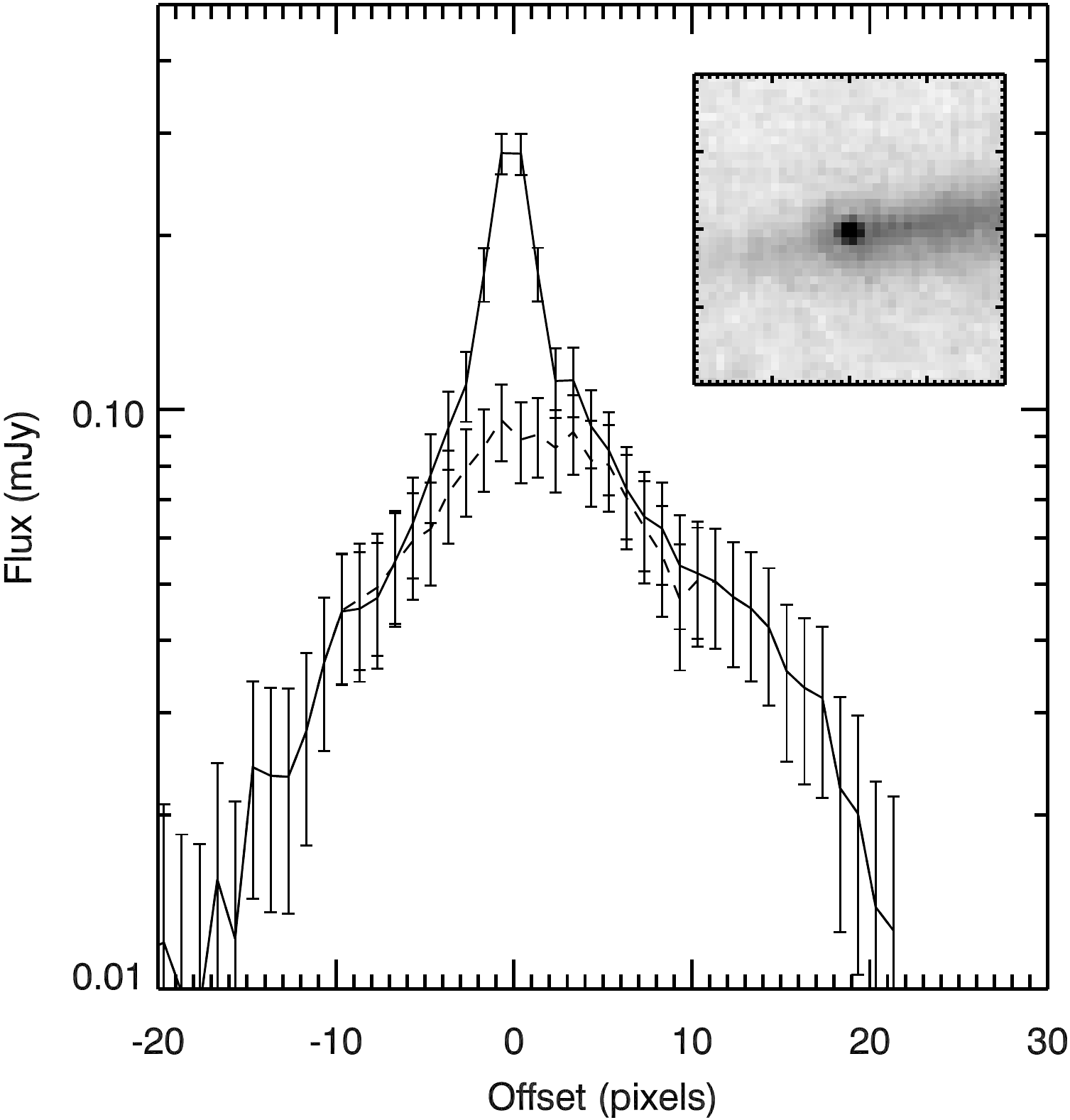}
\caption{Kelley \textit{et al.}, The Dust Trail of Comet 67P/Churyumov-Gerasimenko}
\end{figure}

\begin{figure}
\epsscale{0.6}
\plotone{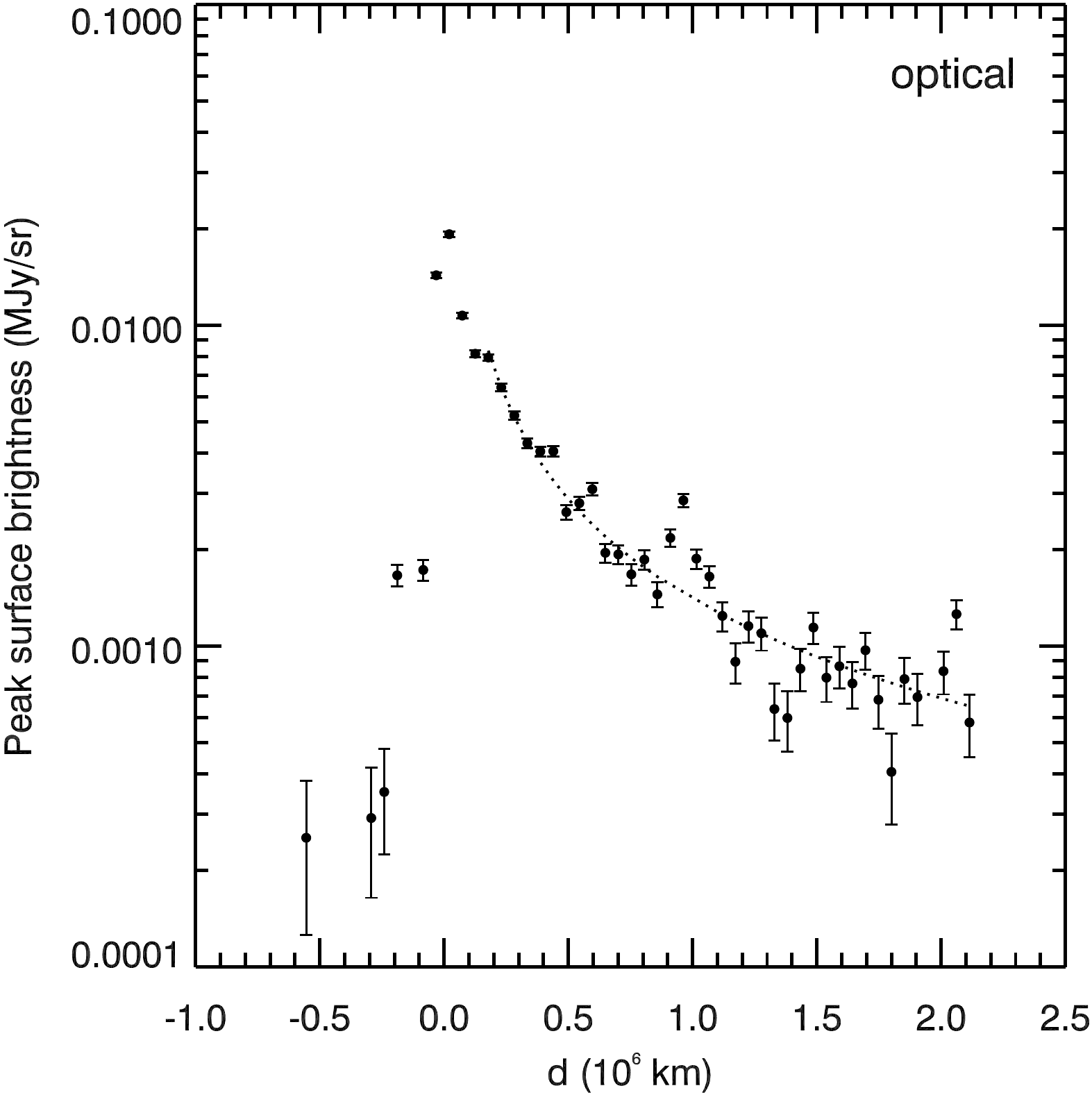}
\plotone{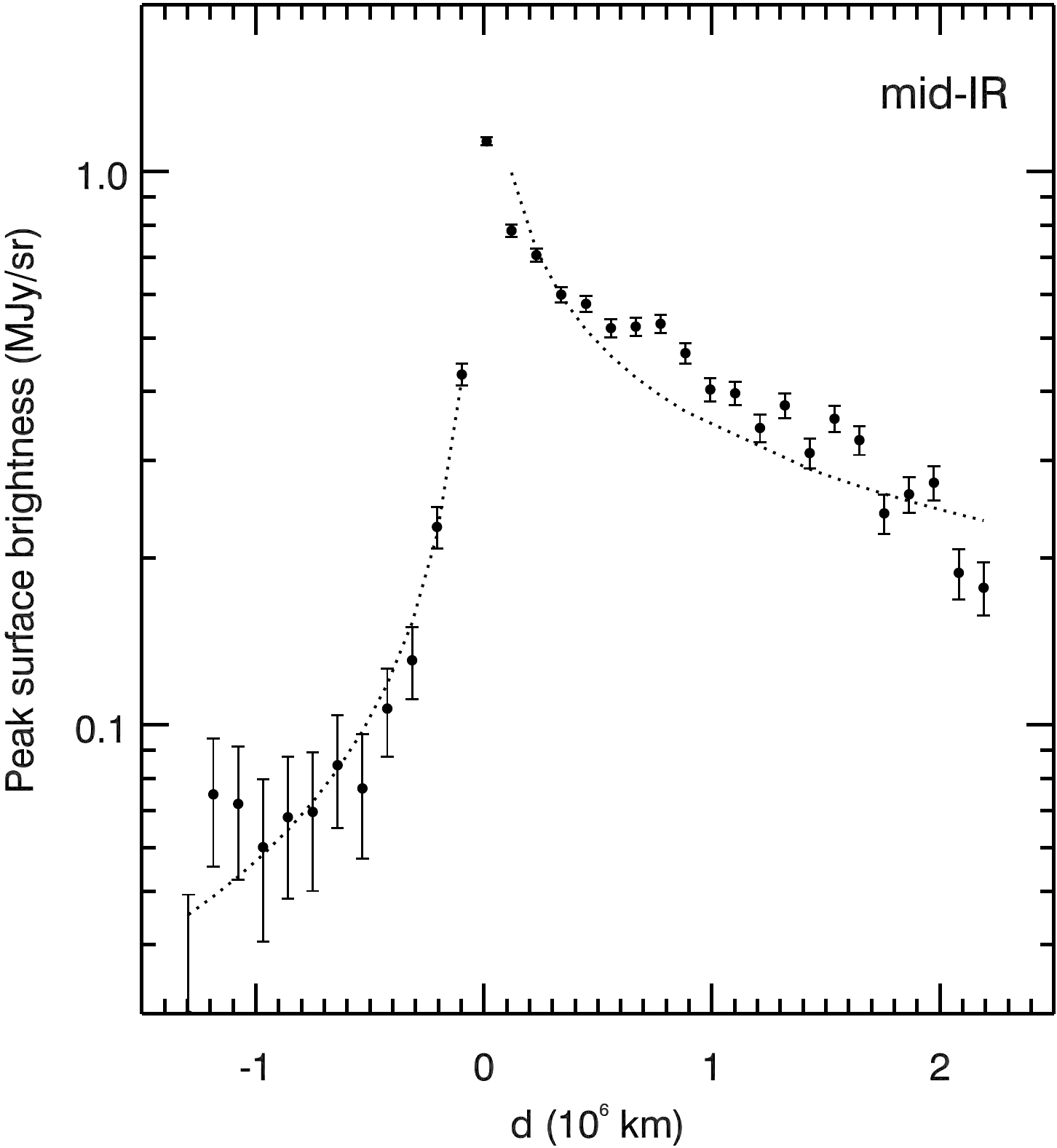}
\epsscale{1}
\caption{Kelley \textit{et al.}, The Dust Trail of Comet 67P/Churyumov-Gerasimenko}
\end{figure}

\begin{figure}
\plotone{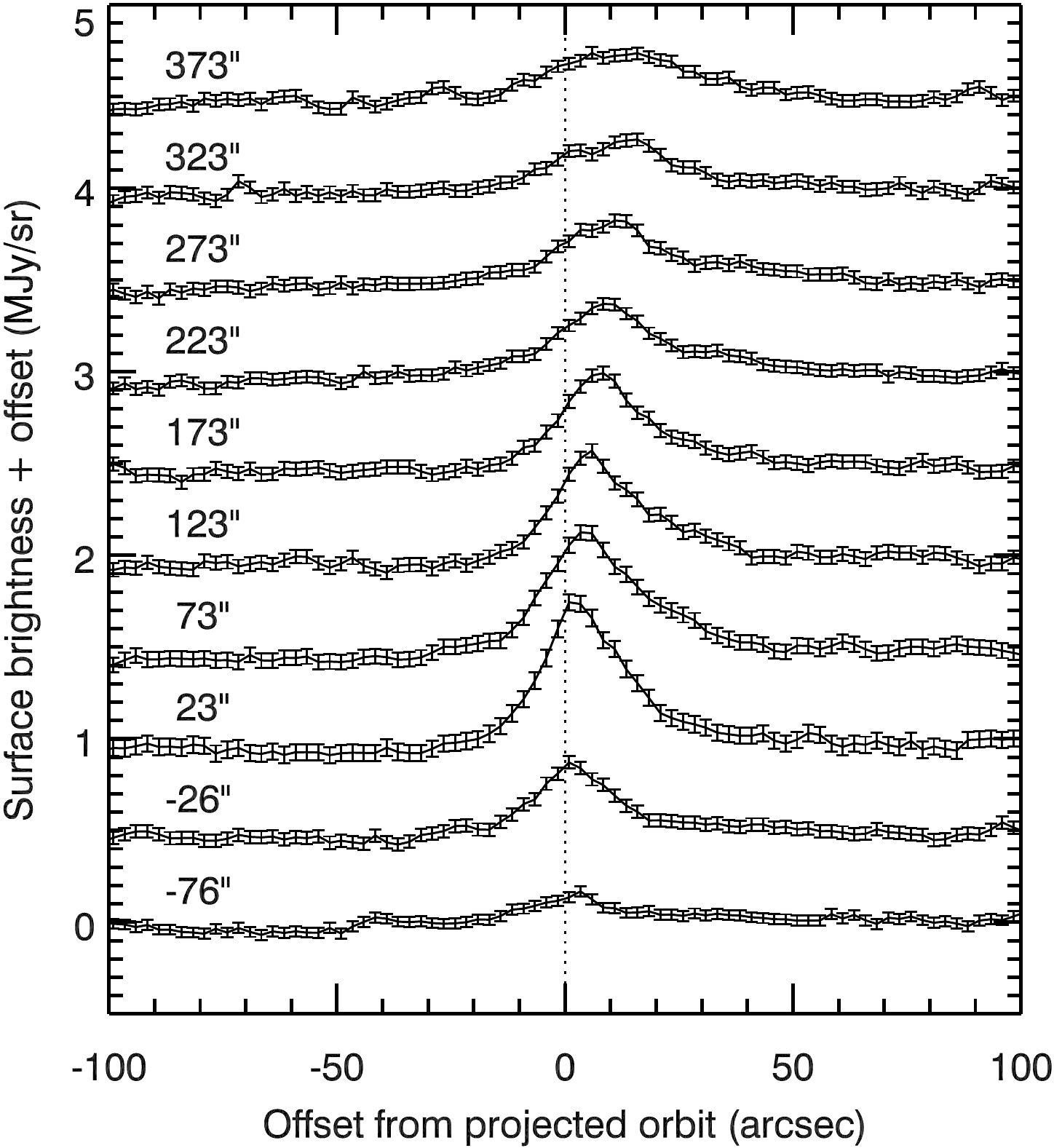}
\caption{Kelley \textit{et al.}, The Dust Trail of Comet 67P/Churyumov-Gerasimenko}
\end{figure}

\begin{figure}
\epsscale{0.59}
\plotone{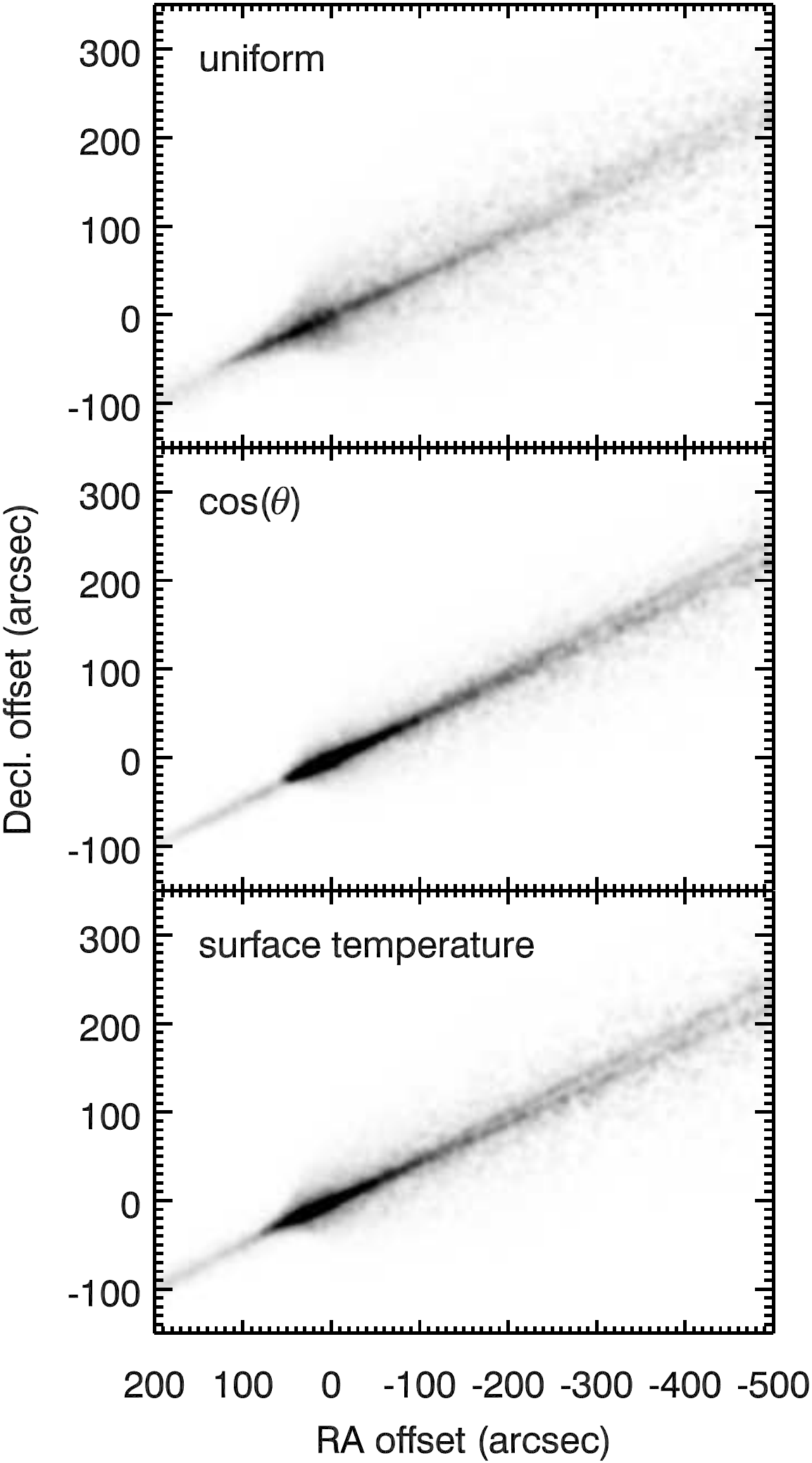}
\epsscale{1}
\caption{Kelley \textit{et al.}, The Dust Trail of Comet 67P/Churyumov-Gerasimenko}
\end{figure}

\begin{figure}
\plotone{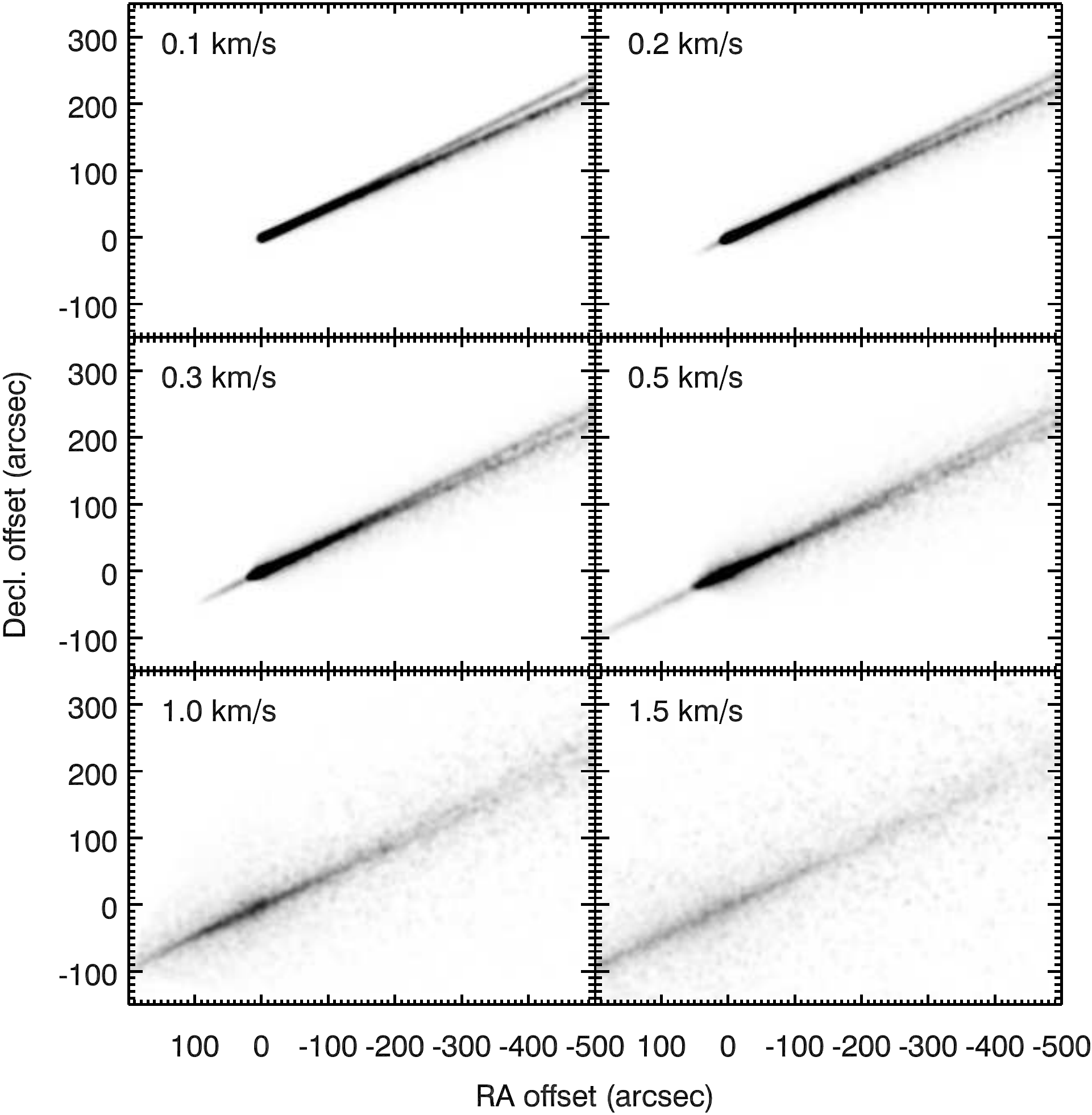}
\caption{Kelley \textit{et al.}, The Dust Trail of Comet 67P/Churyumov-Gerasimenko}
\end{figure}

\begin{figure}
\plotone{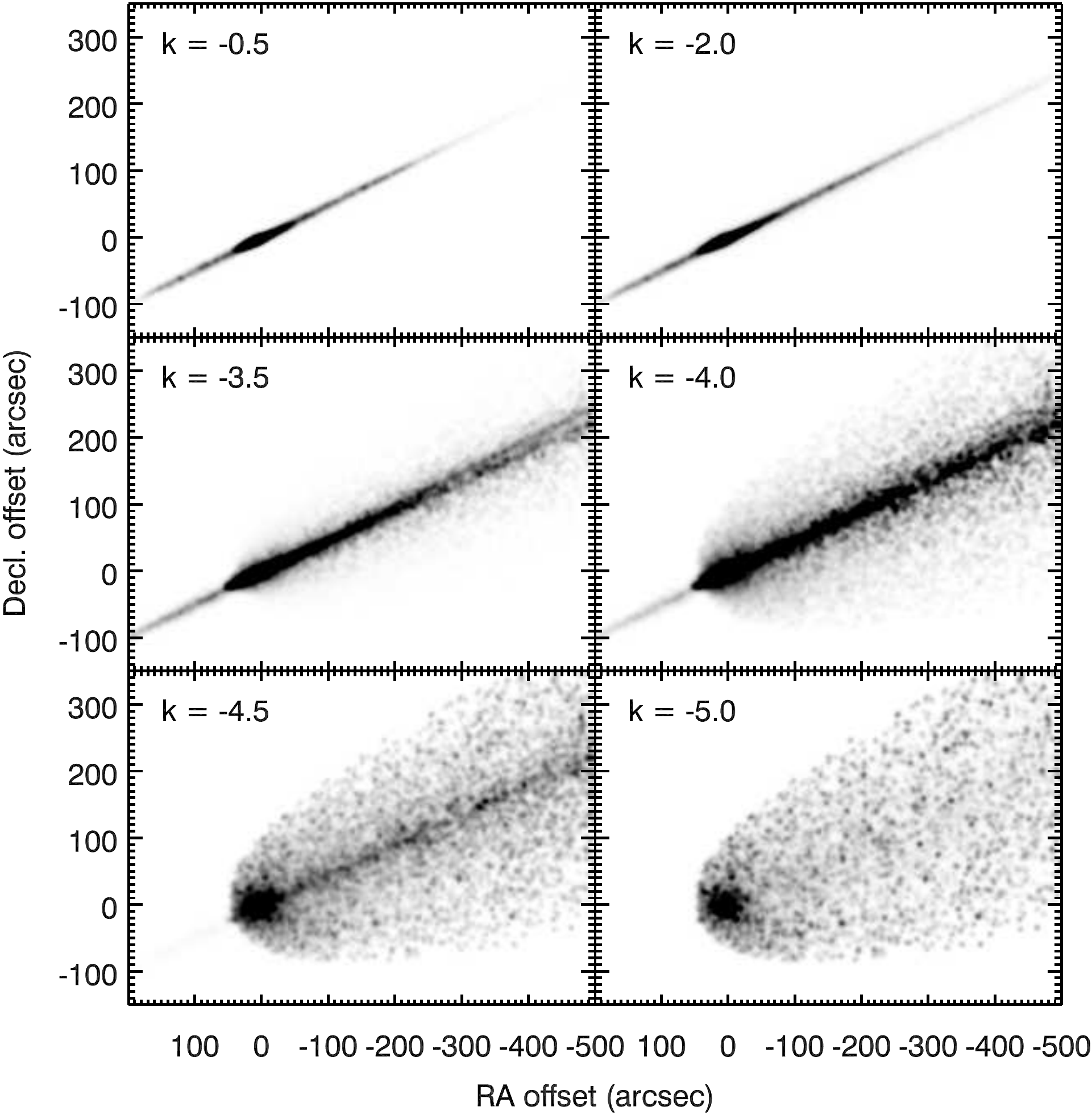}
\caption{Kelley \textit{et al.}, The Dust Trail of Comet 67P/Churyumov-Gerasimenko}
\end{figure}

\begin{figure}
\plotone{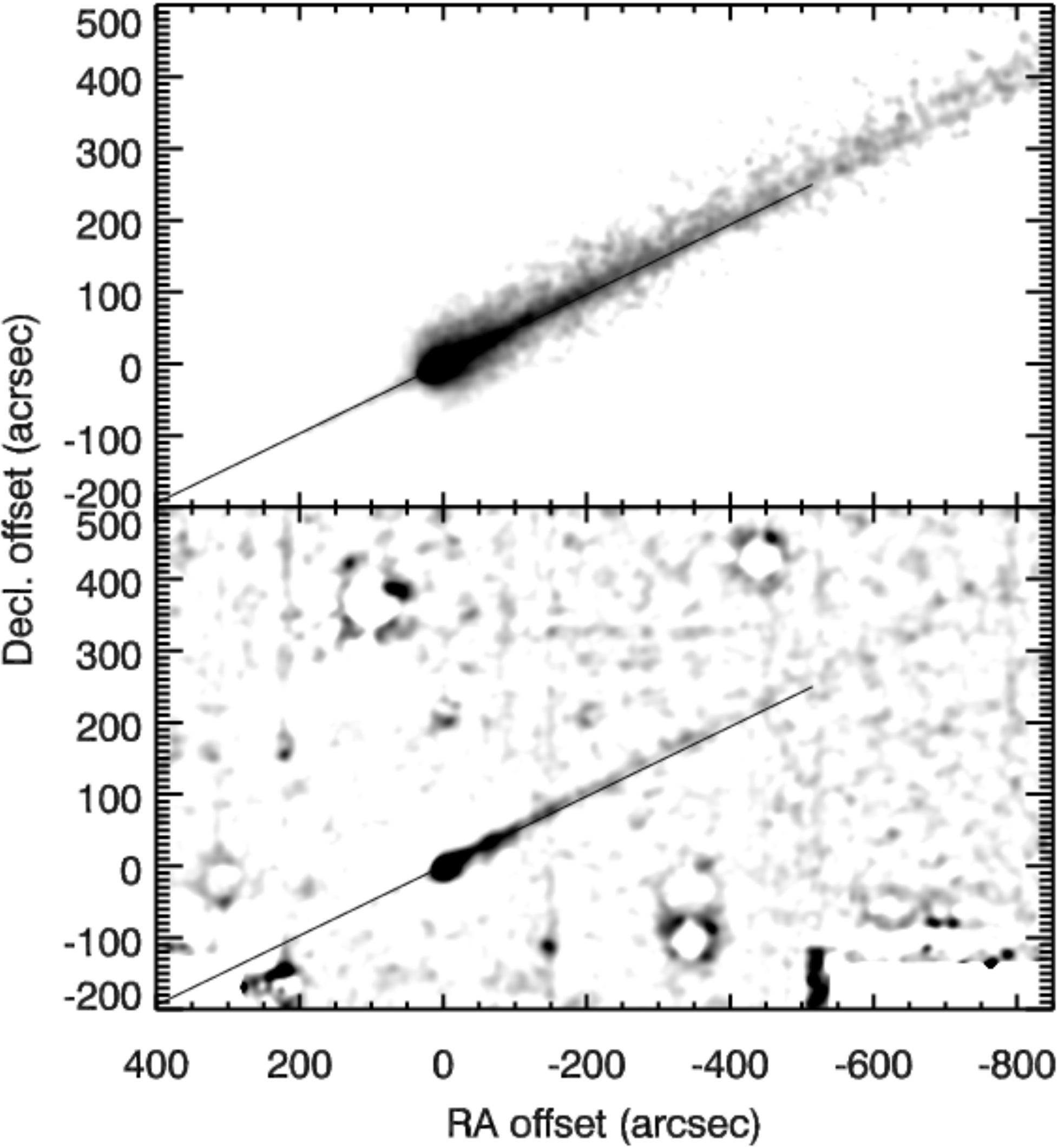}
\caption{Kelley \textit{et al.}, The Dust Trail of Comet 67P/Churyumov-Gerasimenko}
\end{figure}

\begin{figure}
\plotone{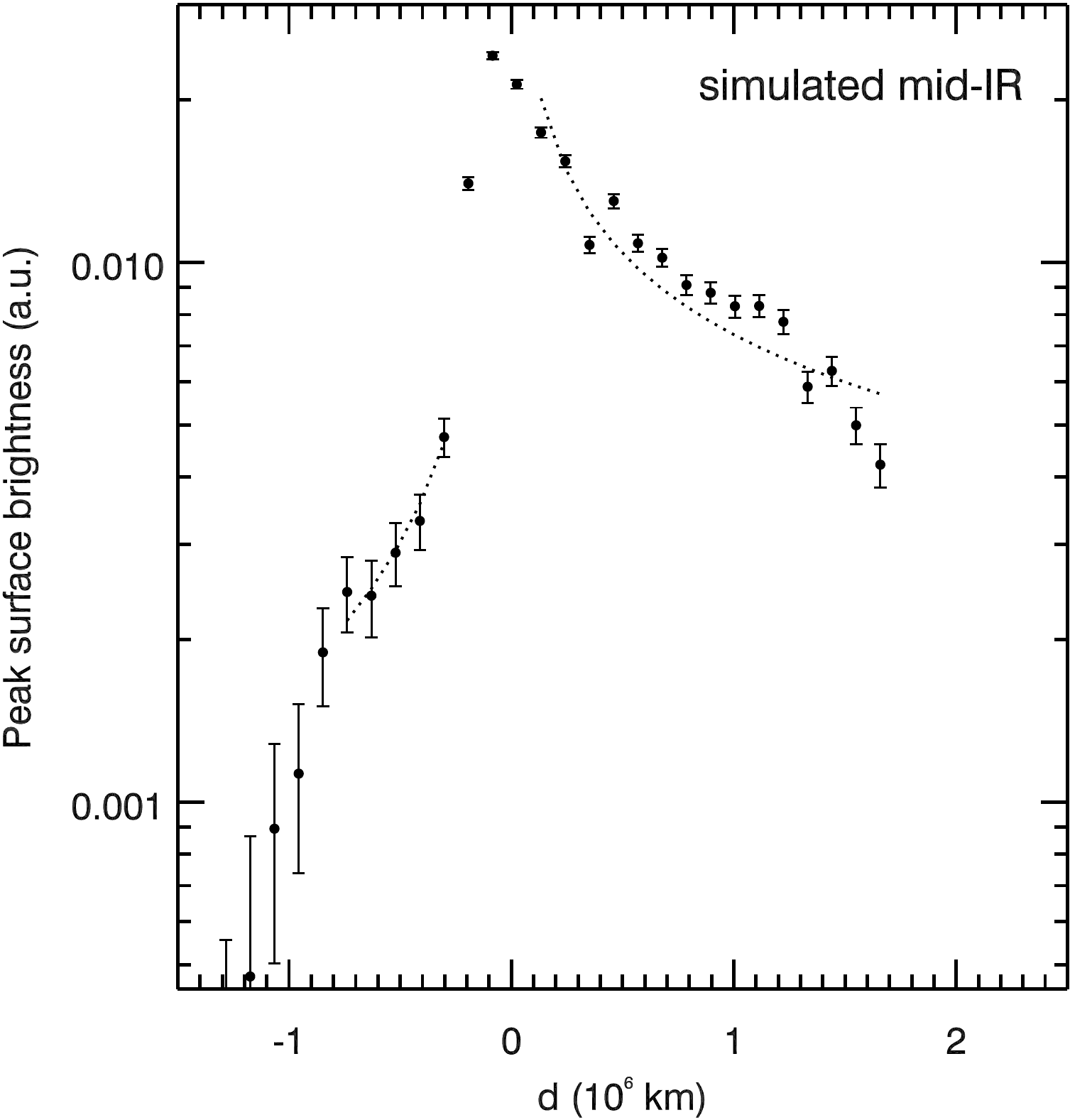}
\caption{Kelley \textit{et al.}, The Dust Trail of Comet 67P/Churyumov-Gerasimenko}
\end{figure}

\begin{figure}
\epsscale{0.59}
\plotone{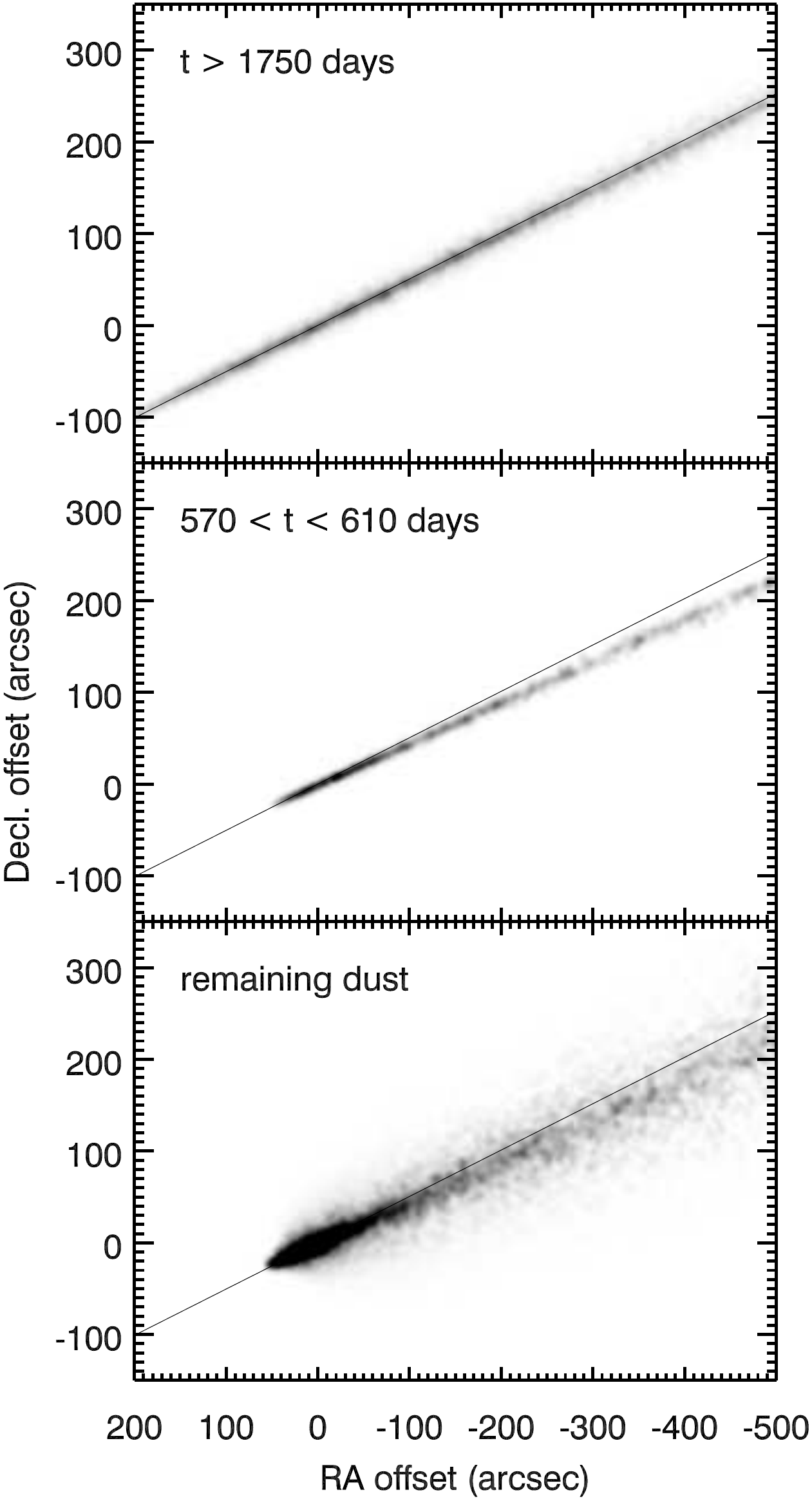}
\epsscale{1}
\caption{Kelley \textit{et al.}, The Dust Trail of Comet 67P/Churyumov-Gerasimenko}
\end{figure}

\end{document}